\documentclass{aa}  
\usepackage{graphicx}
\usepackage{lscape}
\usepackage{txfonts}
\usepackage{natbib}
\usepackage{float}
\usepackage{xcolor}

\begin{document}

\title{Investigation of the WR\,11 field at decimeter wavelengths\thanks{The radio data presented here were obtained with the Giant Metrewave Radio Telescope (GMRT). The GMRT is operated by the National Centre for Radio Astrophysics of the Tata Institute of Fundamental Research.}}

\author{P. Benaglia\inst{1,2}
  \and
       S. del Palacio\inst{1,2}
  \and        
       Ishwara-Chandra, C.H.\inst{3}
  \and
       M. De Becker\inst{4}
  \and
       N. L. Isequilla\inst{2}
  \and
       J. Saponara\inst{1,2}
}

   \institute{Instituto Argentino de Radioastronomia, CONICET \& CICPBA, CC5 (1897) Villa Elisa, Prov. de Buenos Aires, Argentina\\
         \email{paula@iar-conicet.gov.ar}
   \and Facultad de Ciencias Astron\'{o}micas y Geof\'{\i}sicas, UNLP, Paseo del Bosque s/n, 1900, La Plata, Argentina      
   \and National Centre for Radio Astrophysics (NCRA-TIFR), Pune, 411 007, India
   \and Space sciences, Technologies and Astrophysics Research (STAR) Institute, University of Li\`ege, Quartier Agora, 19c, All\'ee du 6 Ao\^ut, B5c, B-4000 Sart Tilman, Belgium\\
}

  \date{Received XXX; accepted YYY}
 
  \abstract 
   {The massive binary system WR\,11 ($\gamma^2$-Velorum) has been recently proposed as the counterpart of a \textit{Fermi} source. If this association is correct, this system would be the second colliding wind binary detected in GeV $\gamma$-rays. However, the reported flux measurements from 1.4 to 8.64~GHz fail to establish the presence of non-thermal (synchrotron) emission from this source. Moreover, WR\,11 is not the only radio source within the \textit{Fermi} detection box. Other possible counterparts have been identified in archival data, some of which present strong non-thermal radio emission. 
   
We conducted arcsec-resolution observations towards WR\,11 at very low frequencies (150 to 1400~MHz) where the non-thermal emission --if existent and not absorbed-- is expected to dominate, and present a catalog of more than 400 radio-emitters, among which a significant part is detected at more than one frequency, including limited spectral index information. Twenty-one of them are located within the \textit{Fermi} significant 
emission. A search for counterparts for this last group pointed at MOST\,0808--471, a source 2' away from WR\,11, as a promising candidate for high-energy emission, with resolved structure along 325 -- 1390~MHz. For it, we reprocessed archive interferometric data up to 22.3~GHz and obtained a non-thermal radio spectral index of $-0.97 \pm 0.09$. However, multiwavelength observations of this source are required to establish its nature and to assess whether it can produce (part of) the observed $\gamma$-rays.

WR\,11 spectrum follows a spectral index of $0.74 \pm 0.03$ from 150~MHz to 230~GHz, consistent with thermal emission. We interpret that any putative synchrotron radiation from the colliding-wind region of this relatively short-period system is absorbed in the photospheres of the individual components. Notwithstanding, the new radio data allowed to derive a mass loss rate of $2.5 \times 10^{-5}$~M$_\sun~{\rm yr}^{-1}$, which, according to the latest  models {for $\gamma$-ray emission in WR 11}, would suffice to provide the required kinetic power to feed non-thermal {radiation} processes.

}

   \maketitle
%
%--------------------------------------------------Sect. 1-----------------

\section{Introduction}
\label{Sec:intro}

Massive stars produce powerful stellar winds. In binary systems, these winds collide, constituting a colliding wind binary (CWB). In the wind collision region (WCR), the shocked winds heat up to several tens of MK. CWBs are expected to radiate throughout the whole electromagnetic spectrum, from radio frequencies to $\gamma$-rays: the theoretical interest of CWBs as candidate $\gamma$-ray sources has been extensively studied in the literature \citep[e.g.,][]{1993ApJ...402..271E,1995IAUS..163..438W,1999A&A...348..868R}.

%-------------------------------------------------------------
%                                      Table: Observing bands
%-------------------------------------------------------------
%
\begin{table*}[ht]
\caption{GMRT observing information and parameters of the images.}  
\label{Tab:1} 
\centering            
\begin{tabular}{l r r r r}  
\hline\hline      
Frequency band centre (MHz) & 150 & 325 & 610 & 1390 \\
Phase calibrator & 0837$-$187 & 0837$-$187 & 0837$-$187 & 0828$-$375 \\
Observing date(s) & 21/Jan, 11/Feb/2017 & 28/May/2016 & 23/Jul/2016 & 03/Oct/2016 \\
Time on source (min) & 241 &171 & 120 & 209 \\
Field of view$\dag$ (arcmin) &  186$\pm$6 & 81$\pm$4 & 43$\pm$3 & 24$\pm$2\\
Synthesized beam & $59.6'' \times 14.0'', 2.5^\circ$ & $22.6'' \times 6.4'' , 8.2^\circ$ & $12.1'' \times 3.2'', 12.5^\circ$ & $5.4'' \times 2.1'', -175.9^\circ$\\
Image centre r.m.s. (mJy beam$^{-1}$) & 1 & 0.2 & 0.16 & 0.07 \\
\hline                     
\end{tabular}
\tablefoot{$\dag$: Taken from the GMRT Observer's Manual; www.ncra.tifr.res.in/ncra/gmrt/gmrt$-$users/observing-help/manual$_-$7jul15.pdf}
\end{table*}
%
%--------------------------------------------------------------------

So far, more than 40 systems have been identified to be relativistic particle accelerators \citep{2013A&A...558A..28D}, thanks to the evidence for non-thermal (NT) radiation found mainly in the radio domain. Such a radio emission is interpreted as synchrotron emission produced by a population of relativistic electrons. These relativistic particles are presumably accelerated in the shock fronts of the WCR by the diffusive shock acceleration (DSA) mechanism \citep{2006ApJ...644.1118R,2006MNRAS.372..801P}.
In general, the radio emission from CWBs is potentially made of three contributions: one thermal, from the individual ionized stellar winds; another one thermal, from the shocked gas in the WCR; and one NT, from the relativistic electrons accelerated in the WCR. The radio emission from a single stellar wind is expected to be steady and to present a flux density $S_\nu \propto \nu^\alpha$, where $\nu$ is the frequency and the spectral index $\alpha$ is close to 0.6 \citep{1975MNRAS.170...41W,1975A&A....39....1P}. The NT radio emission from the WCR at low frequencies is notably higher than that expected for the thermal emission from individual stellar winds. Its spectral index is basically negative, and the flux density and spectral index may vary as a function of time. The thermal free-free emission from the WCR is also modulated with the orbital phase, both in flux density and spectral index. The latter can be close to 1 for the case of radiative shocks \citep{2010MNRAS.403.1633P}.

The only confirmed $\gamma$-ray emitting CWB is $\eta$-Car, as the high-energy $\gamma$-ray emission from this system presents a modulation corresponding to the orbital period of the binary \citep{2009ApJ...698L.142T,2015A&A...577A.100R}. Additionally, $\eta$-Car has been recently reported as a non-thermal hard X-ray source \citep{2018NatAs...2..731H} and a very high-energy $\gamma$-ray source \citep{Leser2017}.

\citet{2016MNRAS.457L..99P} proposed WR\,11 as the second CWB with a $\gamma$-ray counterpart after analyzing almost seven years of Pass-8 \textit{Fermi} data. He discovered a \textit{Fermi} excess at a position coincident to this source with a flux of $(2.7 \pm 0.5) \times 10^{-12}$~erg~cm$^{-2}$~s$^{-1}$. 

In general, the detection of $\gamma$-ray emission implies the presence of relativistic particles. If at least part of those particles are electrons, one could expect to find also the radio synchrotron emission they produce by interacting with ambient magnetic fields.

Archival radio emission from WR~11 presents a spectral index consistent with thermal bremsstrahlung. Nonetheless, this is not sufficient to rule out WR\,11 as the counterpart of the \textit{Fermi} source, as it is most likely that the synchrotron emission from the system --if existent-- would be absorbed by the stellar winds material. If it were possible to detect $\gamma$-ray emission modulated with the period of WR\,11, then it would be a strong evidence supporting that this is the actual counterpart. However, the low high-energy flux does not allow for such timing analysis as the detected excess is found only after integrating during several years.

On the other hand, \citet{2016PASA...33...17B} studied the field of WR\,11 by means of archive Australia Telescope Compact Array (ATCA) data and found several NT radio sources in the \textit{Fermi} detection box that, in principle, could also be related to the counterpart of the \textit{Fermi} source. In this work we carry out an investigation of the WR~11 field-of-view sources below 1.4~GHz in order to analyze which of them are clearly NT particle accelerators, and to infer if any of them could account for the \textit{Fermi} excess. Section~2 summarizes what is known about WR\,11 and its surroundings, relevant to this study. In Sect.~3 we describe the observations and the images generated from them. Section~4 presents the results obtained from the images and some analysis. A discussion mainly of the NT contributions to the radio emission is given in Sect.~5, and Sect.~6 closes with the conclusions.
 
%-----------------------------------------------------Sect. 2---------------

\section{WR\,11 and its surroundings}
\label{Sec:wr11}

Gamma$^2$-Velorum or WR\,11, ($RA, Dec_{\rm J2000}$ = 8:9:31.95, $-$47:20:11.71), is the nearest stellar massive binary system, at a distance of 340~pc. It is composed by a WC8 and an O7.5 star. The orbit semimajor axis is 1.2~AU and the system period is 79~days \citep{2007MNRAS.380.1276N}, thus prone to host a wind-collision region. \cite{2007ASPC..367..159V}, analyzing \textit{XMM-Newton} data, derived a mass loss rate $\dot{M} = 8 \times 10^{-5}$~M$_\odot$~yr$^{-1}$, and detected X-rays dominated by thermal emission from the WCR.

A recent work by \citet{2016MNRAS.457L..99P} claimed that WR\,11 is the counterpart of a \textit{Fermi} 6.1$\sigma$ flux excess, which would make it the second colliding wind binary to be detected in GeV $\gamma$ rays. The author derived a flux in the high-energy range (0.1-- 100~GeV) of $1.8 \pm 0.6 \times 10^{-9}$~ph~cm$^{-2}$~s$^{-1}$ and an energy flux of $2.7 \pm 0.5 \times 10^{-12}$~erg~cm$^{-2}$~s$^{-1}$. At a distance of 340~pc, the flux translates into a luminosity $L = 3.7 \pm 0.7 \times 10^{31}$~erg~s$^{-1}$, which is well below the available wind kinetic luminosity. Numerical models by \cite{2017ApJ...847...40R} support the association of WR\,11 with the $\textit{Fermi}$ source.

The NT nature of WR\,11 radiation is still uncertain. It was detected with ATCA at the lowest frequency of 1.4~GHz with a flux of 9~mJy \citep[see][and references therein]{1999ApJ...518..890C}. The spectral indices derived from 1.4 to 8.64~GHz resulted from 1.2 to 0.3. 
However, \citet{2016PASA...33...17B} reduced archive ATCA data of three projects (identified as C599, C787 and C1616) and found very consistent results for WR\,11. The emission spectral index between 1.4 and 2.4~GHz is equal to $0.85 \pm 0.1$, regardless of the orbital phase. Multi-configuration C787 data averaged along 180~days (from June to December, 2001) allowed to map the field of view of WR 11 and reveal many nearby NT sources. Besides WR~11, seven sources were detected above a nominal threshold of 10~mJy, all of which presented negative spectral indices. In particular, the so-called S6, a double source with intense NT radiation, is located at only 2’ from WR\,11, inside the 6.1$\sigma$ probability contour of the \textit{Fermi} excess.

We observed WR\,11 and its surroundings with the Giant Metrewave Radio Telescope (GMRT) at decimeter wavelengths to investigate the emission regime of the CWB and to address whether other sources in the field could be responsible or contribute to the \textit{Fermi} source.

%-----------------------------------------------------Sect. 3---------------
\section{Radio observations and data reduction}
\label{Sec:observations}

\subsection{GMRT data}

Dedicated observations pointing at WR\,11 were carried out with the GMRT in four frequency bands, centred at 150, 325, 610 and 1390~MHz, using the back end non-polarimetric configuration of 32-MHz bandwidth; see Table~\ref{Tab:1} with the observing details (project code 30\_033). The source had been observed with the ATCA at 1400~MHz \citep[][]{1999ApJ...518..890C}, and we collected data at this band with the GMRT to compare both results. This allowed us to use datasets of the two radio interferometers in a common analysis. Besides, the GMRT angular resolution was $\sim$2 -- 5 times better than the ATCA one.

The source was observed during 2016 on 28~May (325~MHz, 6~h), 23~Jul (610~MHz, 4~h), and 3~Oct (1390~MHz, 4~h), and in 2017 on 21~Jan and 11 Feb (150~MHz, 8~h). The data reduction and imaging were performed with the Source Peeling and Atmospheric Modeling (SPAM) module \citep{2014ASInC..13..469I}, a python-based extension to the Astronomical Imaging Processing System package \citep[AIPS, ][]{2003ASSL..285..109G}, for the 150, 325 and 610~MHz band datasets. We used the Common Astronomy Software Applications \citep[CASA,][]{2007ASPC..376..127M} to  process the 1390-MHz raw data in the standard way and obtained a robust-weighted self-calibrated image. The sources 3C147 and 3C286 were used as primary calibrators. The phase calibrators were 0828--375 at 1390~MHz and 0837--198 for the rest of the bands. The images were built with robust weightings.

The {\sc miriad} software package \citep{1995ASPC...77..433S} and the routine {\sc kvis} of the {\sc karma} visualization software \citep{1996ASPC..101...80G} were used for data analysis. The synthesized beams and the r.m.s. attained at the field center are given in Table \ref{Tab:1}.

\subsection{Archive data}

To complement the investigation, we made use of the Australia Telescope Online Archive (ATOA, https://atoa.atnf.csiro.au/) observations that targeted WR\,11, projects C787 and C1616. All datasets were processed with the {\sc miriad} package in a standard way, obtaining robust-weighted images. The C787 reprocessed data consisted of 0.75~h on source at 4.8~GHz and 3.79~h at 8.64~GHz in 6A array configuration (observations carried out in Dec 23 2001 and May 23 1999, respectively). 
The C1616-6A datasets comprise 4.5~h on-source time at 22.231~GHz, and 4.55~h at 22.367~GHz, imaged together.

%--------------------------------------------------Sect. 4-----------------
\section{Results}
\label{Sec:results}
\subsection{The WR~11 system}

%-------------------------------------------------------------
%                                    Figures 1-2: GMRT images 
%-------------------------------------------------------------
 \begin{figure}
   \includegraphics[width=7cm]{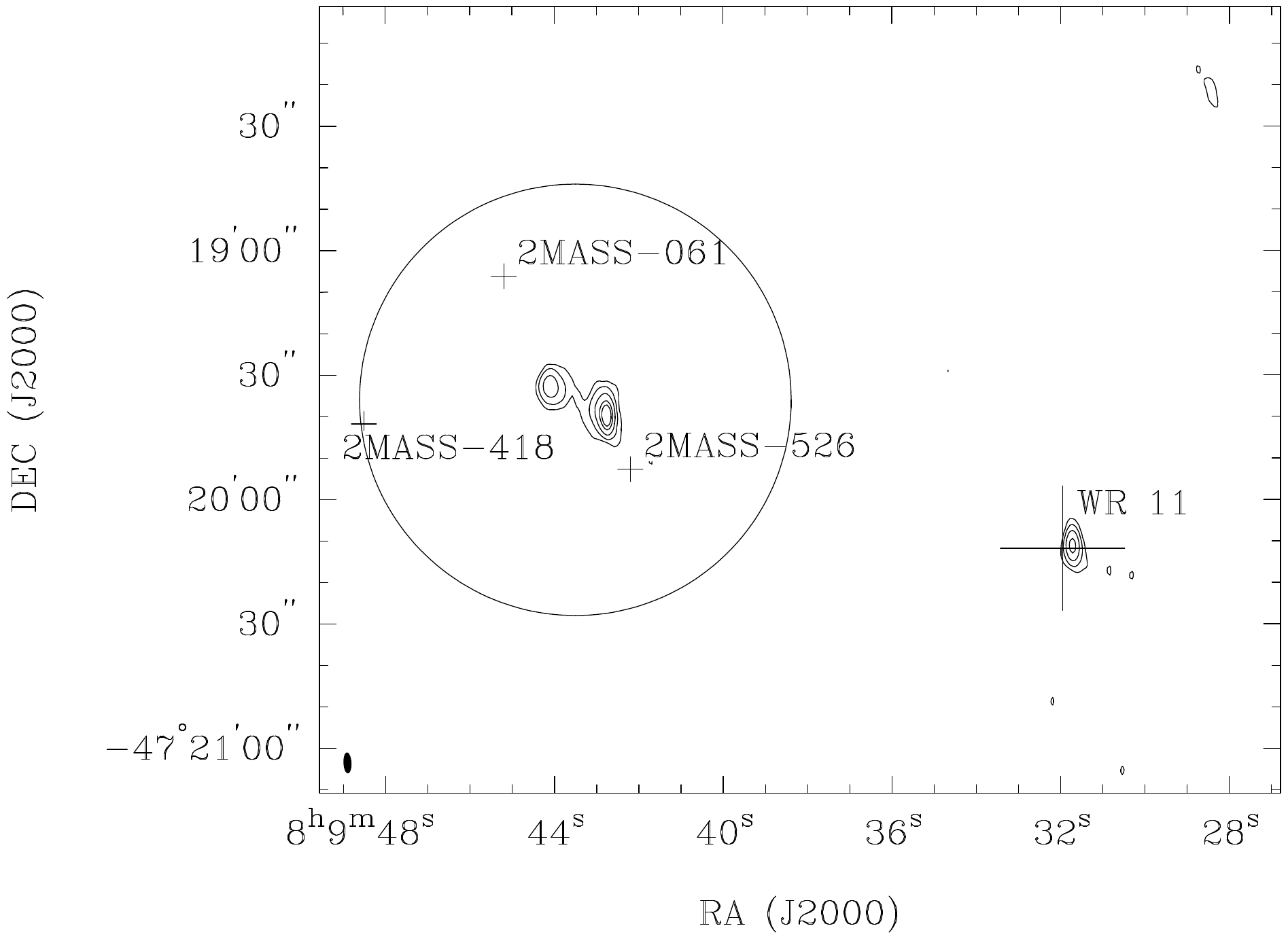}
   \includegraphics[width=7cm]{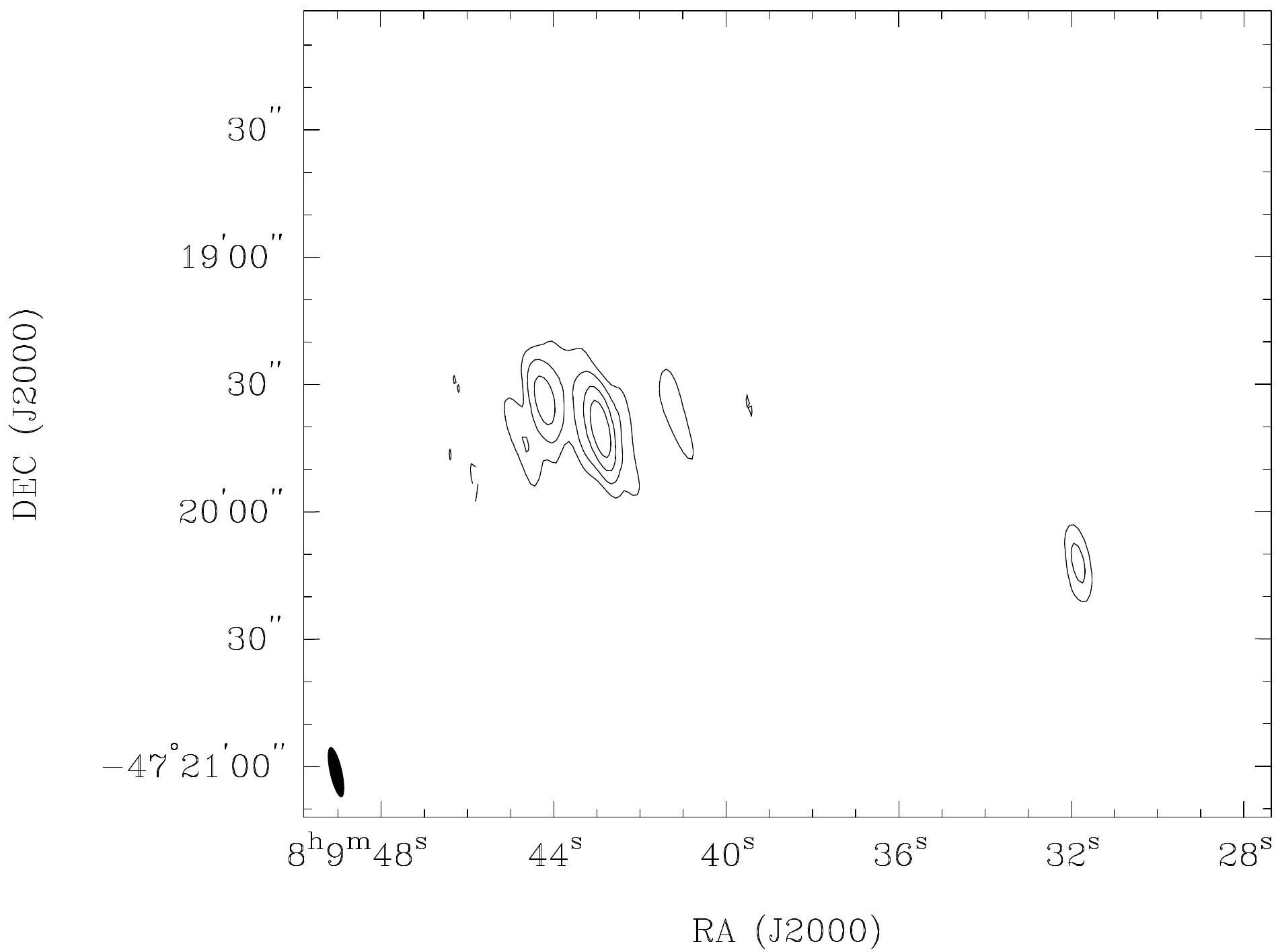}
   \caption{Continuum images of WR\,11 and MOST\,0808--471. Top: at 1390~MHz. Contour levels are $\pm$3, 10, 30, 80 and 130$\sigma$, with $\sigma$ = 0.07~mJy~beam$^{-1}$. The circle represents the beam of the Molonglo Observatory Synthesis Telescope (see text). The sources resulting from a Simbad database search are also marked with crosses. 2MASS-061, 2MASS-418 and 2MASS-526 stand for 2MASS\,J08094519--4719061, 2MASS\,J08094851--4719418 and 2MASS\,J08094219--4719526, respectively. Bottom: At 610~MHz; contour levels of $\pm$3, 10, 30 and 80$\sigma$, with $\sigma$ = 0.16~mJy beam$^{-1}$. Synthesized beams are shown at the bottom left corner of each image.}
    \label{Fig:s5-6a-layout}
 \end{figure}

 \begin{figure}
   \includegraphics[width=7cm]{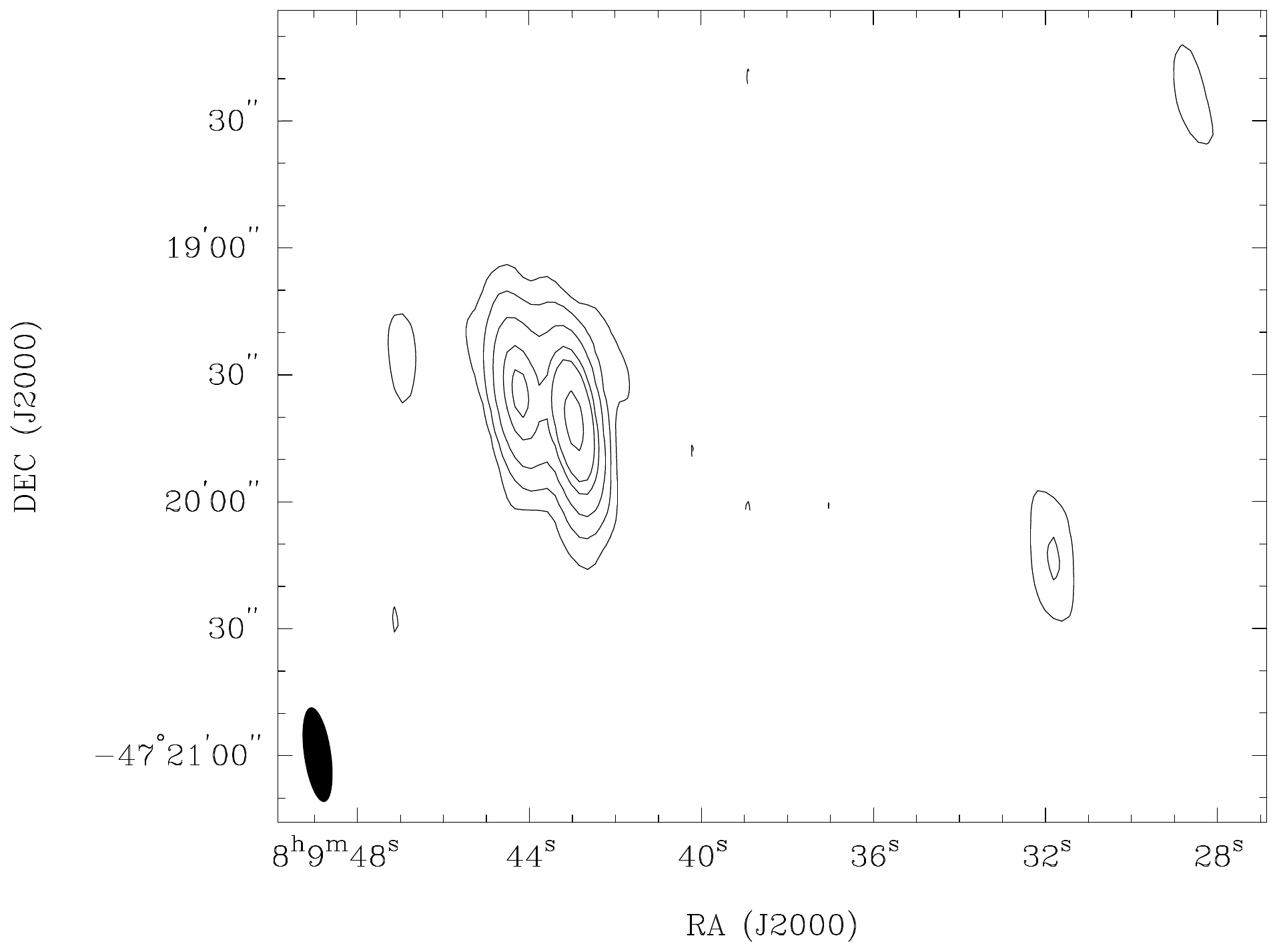}
   \includegraphics[width=7cm]{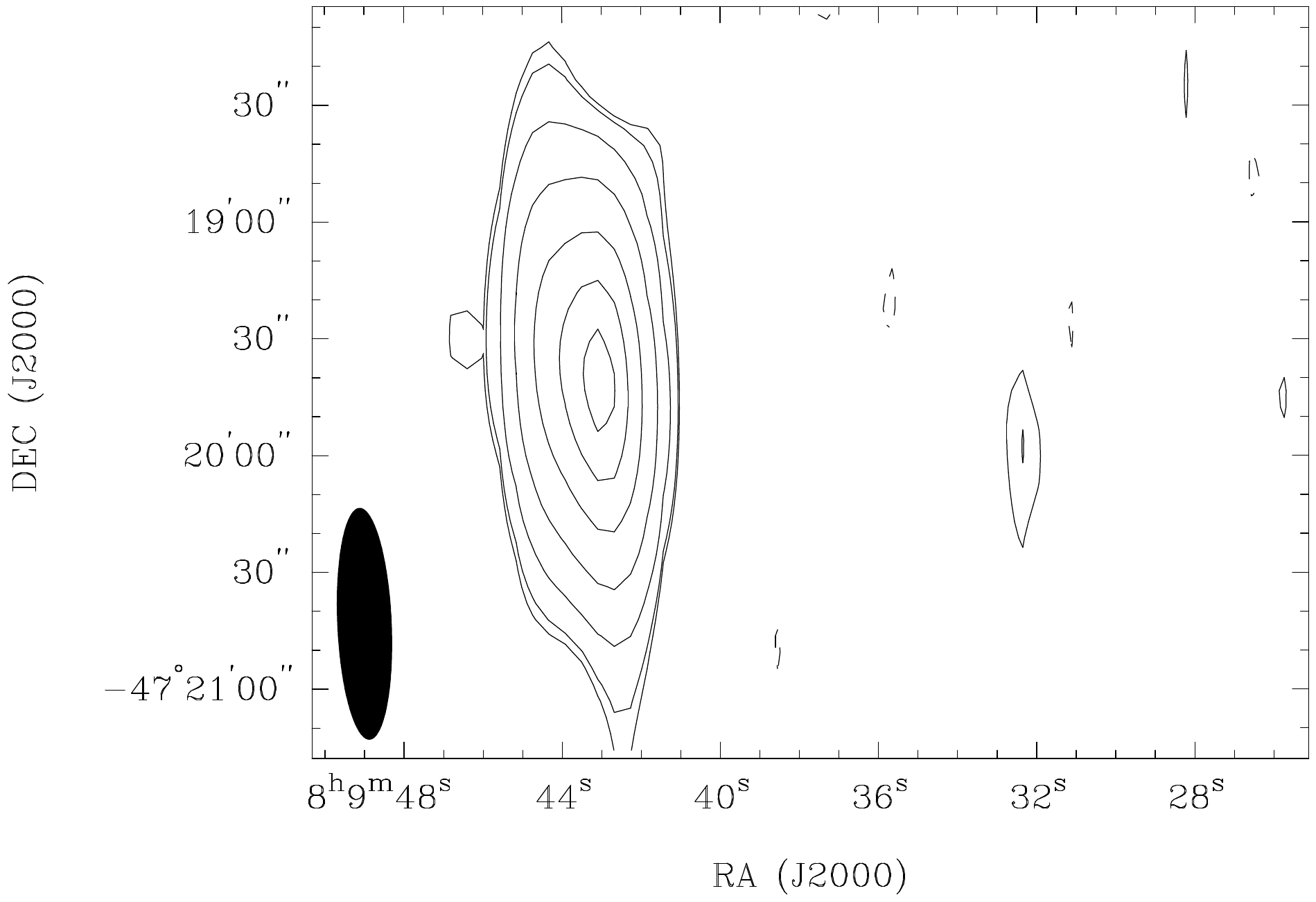}
    \caption{Continuum images of WR\,11 and MOST\,0808--471. Top: At 325~MHz; contour levels of $\pm$3, 10, 30, 80, 130 and 300$\sigma$, with $\sigma$ = 0.2~mJy~beam$^{-1}$. The maxima, from right to left, correspond to sources I23, I25 and I28 of Table~\ref{tab:53sources}, respectively. Bottom: At 150~MHz; contour levels of $\pm$3, 4, 10, 30, 80, 150 and 220$\sigma$, with $\sigma$ = 0.8~mJy~beam$^{-1}$. Synthesized beams are shown at the bottom left corner of each image.}
    \label{Fig:s5-6b-layout}
   \end{figure}
%-------------------------------------------------------------

WR\,11 \citep[source S5 of][]{2016PASA...33...17B} was detected at
325, 610 and 1390~MHz GMRT images. The radio fluxes, obtained by means of a Gaussian fit, resulted in 
$2.9 \pm 0.3$~mJy at 325~MHz, $3.1 \pm 0.2$~mJy at 610~MHz and $10.5\pm1$~mJy at 1390~MHz. At 150~MHz, the rms around the position of WR\,11 is 0.9 mJy~beam$^{-1}$, thus we quote a flux upper limit of 2.7 mJy.
The images are displayed in Figs.~\ref{Fig:s5-6a-layout} \& ~\ref{Fig:s5-6b-layout}. 

Table~\ref{table:2} lists radio fluxes of WR\,11 derived here, together with others gathered from the literature, including the date of the observations. We checked {\sl (i)} the 1.4 to 8.64~GHz fluxes of \citet{1999ApJ...518..890C} by reducing the ATCA archive data already mentioned and {\sl (ii)} the 0.843~GHz one by measuring the flux from the MGPS2 image, obtaining same values among error bars in all cases. 
In particular, we compared the 1.4~GHz flux values taken using three different data sets \citep[][and this work]{1999ApJ...518..890C, 2016PASA...33...17B}. 
The consistency of the results 
supports the study of the spectral energy distribution of WR~11 from 150~MHz to $\sim 22$~GHz combining GMRT and ATCA data. Besides, we found that observations taken with similar array+configuration combinations at different years (1997 and 2001 for 2.4 GHz; 1995 and 2001 for 4.8 and 8.64 GHz; 1997 and 2016 for 1.4 GHz) serve to derive flux values in very good agreement. We conclude that they form a uniform database, and also that they do not show flux variability at the mJy level. In addition, 
the WISE survey catalog \citep{2012yCat.2311....0C} gives the following fluxes of WR\,11: $31.44 \pm 0.5$~Jy at 3.4~$\mu$m, $30.63 \pm 0.66$~Jy at 4.6$~\mu$m, $15.77 \pm 0.15$~Jy at 12~$\mu$m, and $6.90 \pm 0.01$~Jy at 22~$\mu$m. 

\begin{figure}
    \resizebox{\hsize}{!}{
	\includegraphics[width=\hsize, angle=270]{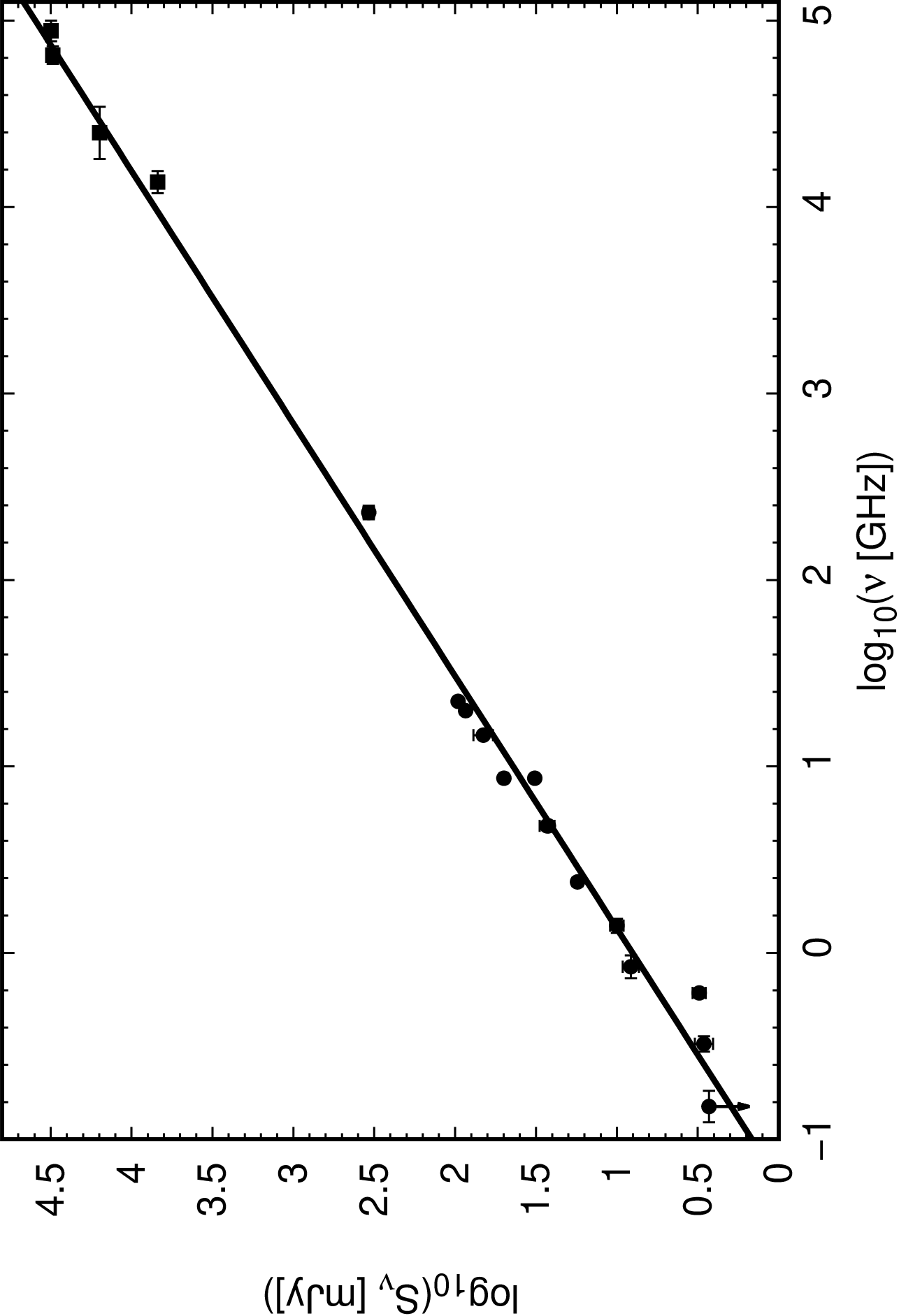}}
    \caption{SED of WR\,11, from radio (bullets) to infrared \citep[squares, WISE catalog,][]{2012yCat.2311....0C} frequencies. The spectral index value of the fit is $+0.74 \pm 0.03$.}
    \label{Fig:SED_WR11}
   \end{figure}

Figure~\ref{Fig:SED_WR11} shows the spectral energy distribution (SED) of WR\,11 from radio to infrared (IR) ranges. The SED is well fitted by a power-law with spectral index \textbf{$\alpha = 0.74 \pm 0.03$},
which is consistent with thermal free-free emission from the ionized stellar winds. A spectral index of 0.6 is the canonical value for individual stellar winds, but an index close to 0.7 might be explained in terms of the influence of acceleration and deceleration zones and/or changes of the ionization structure in the winds \citep[e.g.,][]{1991ApJ...377..629L}. We note that the WCR also produces thermal radio emission that might contribute to the observed fluxes\footnote{The WCR can also produce NT radio emission with its negative spectral index; however, such emission is doomed to be absorbed in the stellar winds for such compact systems.}. An expected index of 1.1 was obtained by \citet{2010MNRAS.403.1633P} and \citet{2011A&A...531A..52M} through numerical calculations of CWBs when the WCR emission dominates, and \citet{2009ApJ...705..899M} observed radio emission with spectral indices $\sim 1$ in various WR systems. However, the WCR emission is expected to be modulated with the orbital phase \citep[e.g.,][]{2010MNRAS.403.1633P}, and the good fit shown in Fig.~\ref{Fig:SED_WR11} for various observations in different epochs points to a rather steady emission. In addition, one should expect the likely thermal emission from the WCR to be significantly absorbed by the dense WC wind material, preventing its signature to be measured or even noticed. Therefore, we conclude that the observed SED from radio to IR in WR\,11 is dominated by the individual winds and is not likely to be significantly contaminated with any emission (either thermal or NT) from the WCR structure.

%-------------------------------------------------------------
%                                    Table: WR 11 radio fluxes
%-------------------------------------------------------------
\begin{table}
\caption{Flux of the WR\,11 system along the radio band.}  
\label{table:2} 
\centering            
%\begin{tabular}{c c c c c}  
\begin{tabular}{c@{~~~}r@{~~~}c@{~~~}c@{~~~}c}
\hline\hline      
Frequency & Flux & Synth. beam & Obs. & Ref. \\
 (GHz) & (mJy) & (arcsec$^2$) & date & \\
\hline       
   0.150 & $<$2.7 & $59.6 \times 14.0$ & 2017.09 & 1 \\ 
   0.325 & $2.9 \pm 0.3$ & $22.6 \times 6.4$  & 2016.41 &1 \\ 
   0.610 & $3.1 \pm 0.2$ & $12.1 \times 3.2$ & 2016.56 &1 \\ 
   0.843 & $8.2 \pm 1$ & $43 \times 43$ & 1983-84 &2 \\ 
   1.4 & $10 \pm 1$ & $6.3 \times 2.3$  & 2016.67 & 1 \\ 
   1.4 & $9.23 \pm 0.13 $ & $12 \times 8$  &   1997.15 & 3 \\  
   2.4 & $17.5 \pm 0.1$ &  $8 \times 5$ & 1997.15  & 3 \\ 
   4.8 & $26.5 \pm 0.28$ & $\sim3 \times 3$ & 1995.49 & 4 \\ 
   8.64 & $32.2 \pm 0.63$ & $\sim1 \times 1$  & 1995.49 & 4 \\  
   14.7 & $67 \pm 10$ & $2.3 \times 2.3$ & 1977.47 & 5 \\ 
   19.9 & $86 \pm 4$ & $10.8 \times 10.8$ & 2004.73 & 6 \\ 
   22.3 & $96 \pm 3$ &  $0.77 \times 0.36$ & 2006.92 & 1 \\ 
   230 & $342 \pm 27$ & $25 \times 25$ & 1990.70 & 7 \\ 
\hline                     
\end{tabular}
\tablebib{(1)~This work; (2)~\citet{1985MNRAS.216..613J}; (3)~\citet{1999ApJ...518..890C}; (4)~\citet{1997ApJ...481..898L};
(5)~\citet{1978MNRAS.182P..47M}); (6)~\citet{2009yCat..74022403M} (AT20G);  (7)~\citet{1991ApJ...377..629L}.}
\end{table}
%-------------------------------------------------------------

%-------------------------------------------------------------
%                                      Figure 3: WR 11 SED
%-------------------------------------------------------------
%-------------------------------------------------------------

%-------------------------------------------------------------
%                                         MOST 0808 radio fluxes
%-------------------------------------------------------------
\begin{table}[h]
\caption{Radio fluxes of the source MOST\,0808--471 up to 9~GHz.}  
\label{Tab:most-table} 
\centering            
\begin{tabular}{l r c c}  
\hline\hline      
Frequency & Flux & Reference & Structure  \\
 (GHz) & (mJy) & & \\
\hline       
   0.150 & $330 \pm 10$ & this work & unresolved \\  
   0.325 & $140 \pm 1$ & this work & resolved \\ 
   \,\,\,\,\,\, left lobe& $44.6 \pm 0.22$ &  &  \\  
   \,\,\,\,\,\, right lobe& $96.3 \pm 0.23$ &  &  \\ 
   0.610 & $66 \pm 3$ & this work &  resolved \\ 
   \,\,\,\,\,\, left lobe & $17.3 \pm 0.3$ &  &  \\ 
   \,\,\,\,\,\, right lobe& $50.4 \pm 0.2$ &  &  \\ 
   0.843 & $69 \pm 2$ & 1 & unresolved \\ 
   1.39 & $39 \pm 3$ & this work & resolved \\ 
   \,\,\,\,\,\, left lobe& $8.7 \pm 0.6$ &  &  \\
   \,\,\,\,\,\, right lobe& $24.6 \pm 1.6$ &  &  \\
   2.4 & $25 \pm 2$ & 2 & unresolved \\  
   4.8 & $6.5 \pm 1$ & this work & unresolved \\  
   8.64 & $2.0 \pm 0.3$ & this work & unresolved \\  
\hline                     
\end{tabular}
\tablebib{(1)~\citet{1985MNRAS.216..613J};
(2)~\citet{2016PASA...33...17B}.
}
\end{table}

%-------------

\subsection{MOST\,0808--471}
\label{most-source}

Using the Molonglo Observatory Synthesis Telescope (MOST), \citet{1985MNRAS.216..613J} detected a radio source at a distance of 2' from WR\,11 with a flux of $69 \pm 2$~mJy, thereafter MOST~0808$-$471, brighter than WR\,11 at 843~MHz. Thus, any measurement towards WR\,11 performed with a beam larger than 2' might be contaminated with emission from this source (at least at cm wavelengths). In \citet{2016PASA...33...17B} it was named as S6. Due to the close proximity of S6 to WR\,11, and since one of the goals of the present investigation is to reveal whether other source(s) besides WR\,11 could be responsible or contribute to the \textit{Fermi} excess detected by \citet{2016MNRAS.457L..99P}, we provide here a more detailed description of the results for this source. 

The images of MOST\,0808--471 at 150, 325, 610 and 1390~MHz are presented in Figs.~\ref{Fig:s5-6a-layout} \& ~\ref{Fig:s5-6b-layout}. The fluxes of MOST\,0808--471 at 150, 325 and 610 were obtained performing Gaussian fitting to the GMRT images. The value at 1390~MHz, frequency at which the source was resolved, was estimated measuring the flux above 3$\sigma$. The S6 flux as estimated from the MGPS2 survey image cutouts \citep{2007yCat.8082....0M} agrees with that of \citet{1985MNRAS.216..613J}. The 4.8 and 8.64~GHz flux values were derived from the images produced using the archive data mentioned in Sect.~\ref{Sec:observations}.

\begin{figure}
    \resizebox{\hsize}{!}{
	\includegraphics[width=\hsize, angle=270]{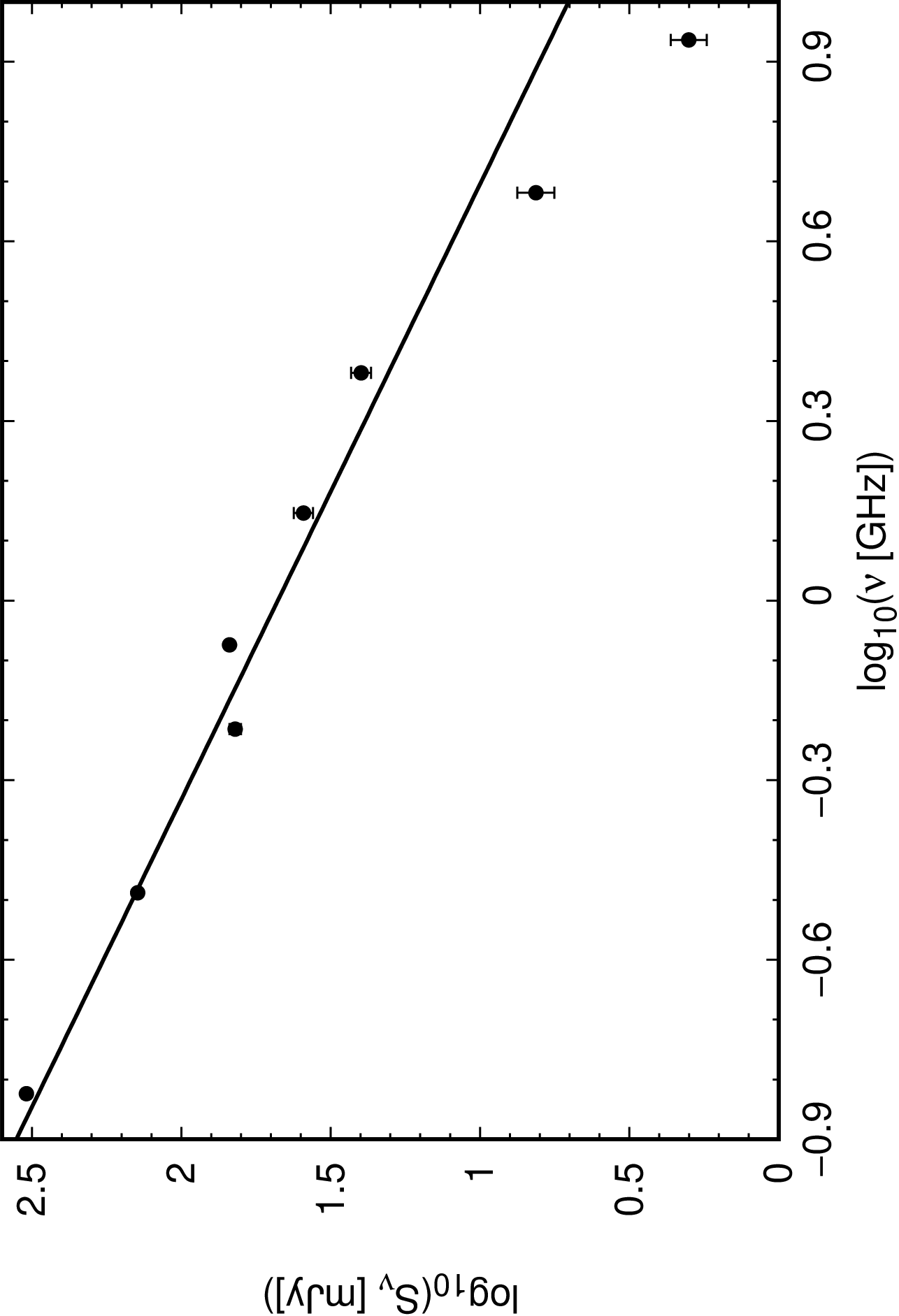}}
    \caption{Radio flux of MOST\,0808--471 as a function of the observing frequency. The spectral index value of the fit is $-0.97 \pm 0.09$.}
    \label{Fig:SED_MOST0808}
   \end{figure}

The series of measurements presented in Table\,\ref{Tab:most-table} allowed us to build the very first radio SED of MOST\,0808--471, shown in Fig.~\ref{Fig:SED_MOST0808}. The plot reveals a clear non-thermal radio spectrum over about two decades of frequencies. A discussion of the nature of this spectrum is presented in Sect.\,\ref{disc-most}.

%-------------------------------------------------------------
%                                      Figure: MOST0808 SED
%-------------------------------------------------------------
%-------------------------------------------------------------

\subsection{Other sources in the field} \label{subsec:other_sources}

%-------------------------------------------------------------
%                      Table of pyBSDF sources: sample records  
%-------------------------------------------------------------

% New table (copy-paste from Jota)
\begin{sidewaystable*}
\caption{Sources detected at 150, 325 and 610~MHz {\bf above the 7$\sigma$ level} (first records); full table will appear as on-line material.}
\label{Tab:pyBSDFshort}
\centering
\begin{tabular}{l r r r r r r r r r} 
\hline\hline    
ID & $RA_{\rm J2000}$ & $Dec_{\rm J2000}$ & Total flux & Peak flux & $RA_{\rm max}$ & $Dec_{\rm max}$ & $B_\mathrm{maj}$ & $B_\mathrm{min}$ & $B_\mathrm{pa}$ \\
   & (hms) & (dms) & (mJy) & (mJy/beam)     &  (hms) &  (dms)  & ($''$)  & ($''$)  & (\degr) \\
\hline
  \multicolumn{10}{c}{\it Detected at 150~MHz}\\ 
\hline
G150-	1	&	8	:	21	:	03.5	$\pm$	0.12	&	-47	:	36	:	05	$\pm$	6.4	&	125.1	$\pm$	8.6	&	61.8	$\pm$	5.0	&	8	:	21	:	03.5	$\pm$	0.12	&	-47	:	36	:	05	$\pm$	6.4	&	100.5	$\pm$	15.6	&	16.9	$\pm$	0.7	&	12.7	$\pm$	1.4	\\
G150-	2	&	8	:	20	:	46.1	$\pm$	0.01	&	-47	:	36	:	07	$\pm$	0.1	&	16622.6	$\pm$	33.5	&	4302.5	$\pm$	6.8	&	8	:	20	:	46.5	$\pm$	0.01	&	-47	:	36	:	11	$\pm$	0.1	&	78.8	$\pm$	0.2	&	41.7	$\pm$	0.1	&	161.3	$\pm$	0.2	\\
G150-	3	&	8	:	20	:	39.8	$\pm$	0.05	&	-47	:	34	:	39	$\pm$	3.5	&	307.0	$\pm$	7.8	&	87.8	$\pm$	6.0	&	8	:	20	:	39.8	$\pm$	0.05	&	-47	:	34	:	39	$\pm$	3.5	&	93.1	$\pm$	8.3	&	31.4	$\pm$	1.7	&	175.1	$\pm$	2.8	\\
G150-	4	&	8	:	20	:	41.4	$\pm$	0.12	&	-47	:	38	:	52	$\pm$	6.3	&	36.6	$\pm$	11.0	&	34.0	$\pm$	5.4	&	8	:	20	:	41.4	$\pm$	0.12	&	-47	:	38	:	52	$\pm$	6.3	&	54.4	$\pm$	15.3	&	16.5	$\pm$	1.5	&	12.8	$\pm$	5.8	\\
G150-	5	&	8	:	20	:	45.2	$\pm$	0.03	&	-47	:	38	:	49	$\pm$	4.3	&	14.6	$\pm$	14.6	&	33.8	$\pm$	3.6	&	8	:	20	:	45.2	$\pm$	0.03	&	-47	:	38	:	49	$\pm$	4.3	&	37.7	$\pm$	10.2	&	9.6	$\pm$	0.4	&	3.4	$\pm$	1.6	\\
G150-	6	&	8	:	20	:	27.7	$\pm$	0.03	&	-46	:	55	:	30	$\pm$	2.0	&	89.8	$\pm$	5.4	&	59.7	$\pm$	3.2	&	8	:	20	:	27.7	$\pm$	0.03	&	-46	:	55	:	30	$\pm$	2.0	&	58.5	$\pm$	4.7	&	21.5	$\pm$	0.8	&	4.5	$\pm$	1.4	\\
G150-	7	&	8	:	20	:	33.1	$\pm$	0.02	&	-47	:	35	:	48	$\pm$	0.1	&	1360.0	$\pm$	11.8	&	691.6	$\pm$	6.2	&	8	:	20	:	34.7	$\pm$	0.02	&	-47	:	35	:	47	$\pm$	0.1	&	57.4	$\pm$	0.7	&	28.6	$\pm$	0.2	&	175.1	$\pm$	0.8	\\
G150-	8	&	8	:	20	:	13.1	$\pm$	0.02	&	-46	:	46	:	45	$\pm$	1.6	&	119.4	$\pm$	4.9	&	70.6	$\pm$	3.1	&	8	:	20	:	13.1	$\pm$	0.02	&	-46	:	46	:	45	$\pm$	1.6	&	59.0	$\pm$	3.7	&	24.0	$\pm$	0.8	&	0.3	$\pm$	1.1	\\
G150-	9	&	8	:	20	:	19.3	$\pm$	0.02	&	-47	:	42	:	52	$\pm$	1.5	&	439.6	$\pm$	7.1	&	172.3	$\pm$	5.1	&	8	:	20	:	19.3	$\pm$	0.02	&	-47	:	42	:	52	$\pm$	1.5	&	82.5	$\pm$	3.4	&	25.9	$\pm$	0.5	&	173.5	$\pm$	179.9	\\
G150-	10	&	8	:	20	:	05.2	$\pm$	0.05	&	-46	:	54	:	59	$\pm$	4.4	&	47.5	$\pm$	4.9	&	27.8	$\pm$	3.1	&	8	:	20	:	05.2	$\pm$	0.05	&	-46	:	54	:	59	$\pm$	4.4	&	63.5	$\pm$	10.5	&	22.5	$\pm$	1.7	&	0.2	$\pm$	5.0	\\
\hline
  \multicolumn{10}{c}{\it Detected at 325~MHz} \\
\hline
G325-	1	&	8	:	14	:	45.5	$\pm$	0.01	&	-47	:	21	:	41	$\pm$	0.1	&	63.9	$\pm$	0.5	&	44.1	$\pm$	0.3	&	8	:	14	:	45.5	$\pm$	0.01	&	-47	:	21	:	41	$\pm$	0.1	&	23.5	$\pm$	0.2	&	8.9	$\pm$	0.0	&	6.1	$\pm$	179.4	\\
G325-	2	&	8	:	14	:	37.4	$\pm$	0.02	&	-47	:	02	:	29	$\pm$	1.2	&	7.7	$\pm$	0.4	&	3.7	$\pm$	0.3	&	8	:	14	:	37.4	$\pm$	0.02	&	-47	:	02	:	29	$\pm$	1.2	&	27.1	$\pm$	3.0	&	11.2	$\pm$	0.7	&	4.5	$\pm$	5.0	\\
G325-	3	&	8	:	14	:	30.3	$\pm$	0.01	&	-47	:	29	:	26	$\pm$	0.1	&	52.1	$\pm$	0.5	&	35.3	$\pm$	0.3	&	8	:	14	:	30.3	$\pm$	0.01	&	-47	:	29	:	26	$\pm$	0.1	&	23.8	$\pm$	0.3	&	9.0	$\pm$	0.1	&	6.4	$\pm$	179.6	\\
G325-	4	&	8	:	14	:	23.6	$\pm$	0.01	&	-47	:	38	:	09	$\pm$	0.7	&	10.9	$\pm$	0.5	&	6.7	$\pm$	0.3	&	8	:	14	:	23.6	$\pm$	0.01	&	-47	:	38	:	09	$\pm$	0.7	&	25.6	$\pm$	1.7	&	9.2	$\pm$	0.3	&	4.4	$\pm$	1.9	\\
G325-	5	&	8	:	13	:	45.4	$\pm$	0.01	&	-46	:	47	:	22	$\pm$	0.7	&	8.6	$\pm$	0.5	&	6.4	$\pm$	0.3	&	8	:	13	:	45.4	$\pm$	0.01	&	-46	:	47	:	22	$\pm$	0.7	&	24.8	$\pm$	1.6	&	7.9	$\pm$	0.2	&	7.4	$\pm$	1.4	\\
G325-	6	&	8	:	13	:	45.7	$\pm$	0.01	&	-47	:	14	:	27	$\pm$	0.1	&	77.8	$\pm$	0.4	&	56.9	$\pm$	0.2	&	8	:	13	:	45.7	$\pm$	0.01	&	-47	:	14	:	27	$\pm$	0.1	&	23.7	$\pm$	0.1	&	8.4	$\pm$	0.0	&	6.8	$\pm$	179.4	\\
G325-	7	&	8	:	13	:	43.0	$\pm$	0.01	&	-47	:	47	:	30	$\pm$	1.2	&	4.7	$\pm$	0.4	&	3.6	$\pm$	0.2	&	8	:	13	:	43.0	$\pm$	0.01	&	-47	:	47	:	30	$\pm$	1.2	&	25.2	$\pm$	2.8	&	7.5	$\pm$	0.3	&	5.1	$\pm$	2.7	\\
G325-	8	&	8	:	13	:	40.3	$\pm$	0.01	&	-47	:	55	:	37	$\pm$	0.1	&	126.6	$\pm$	0.6	&	76.1	$\pm$	0.3	&	8	:	13	:	40.5	$\pm$	0.01	&	-47	:	55	:	36	$\pm$	0.1	&	24.4	$\pm$	0.1	&	9.7	$\pm$	0.0	&	1.8	$\pm$	0.3	\\
G325-	9	&	8   :	13	:	32.3	$\pm$	0.01	&	-47	:	31	:	09	$\pm$	0.2	&	20.2	$\pm$	0.3	&	15.3	$\pm$	0.2	&	8	:	13	:	32.3	$\pm$	0.01	&	-47	:	31	:	09	$\pm$	0.2	&	23.8	$\pm$	0.5	&	8.1	$\pm$	0.1	&	6.6	$\pm$	180.0	\\
G325-	10	&	8	:	13	:	32.7	$\pm$	0.01	&	-47	:	40	:	31	$\pm$	0.7	&	6.3	$\pm$	0.4	&	4.8	$\pm$	0.2	&	8	:	13	:	32.7	$\pm$	0.01	&	-47	:	40	:	31	$\pm$	0.7	&	24.2	$\pm$	1.7	&	8.0	$\pm$	0.2	&	7.2	$\pm$	1.7	\\
\hline
  \multicolumn{10}{c}{\it Detected at 610~MHz} \\
\hline
G610-	1	&	8	:	11	:	46.0	$\pm$	0.01	&	-47	:	26	:	13	$\pm$	0.8	&	2.2	$\pm$	0.2	&	1.4	$\pm$	0.1	&	8	:	11	:	46.0	$\pm$	0.01	&	-47	:	26	:	13	$\pm$	0.8	&	12.7	$\pm$	1.9	&	4.8	$\pm$	0.3	&	7.4	$\pm$	6.1	\\
G610-	2	&	8	:	11	:	39.8	$\pm$	0.02	&	-47	:	15	:	42	$\pm$	1.0	&	1.4	$\pm$	0.2	&	1.1	$\pm$	0.1	&	8	:	11	:	39.8	$\pm$	0.02	&	-47	:	15	:	42	$\pm$	1.0	&	13.6	$\pm$	2.3	&	3.5	$\pm$	0.2	&	14.4	$\pm$	3.8	\\
G610-	3	&	8	:	11	:	17.2	$\pm$	0.01	&	-47	:	25	:	40	$\pm$	0.3	&	3.7	$\pm$	0.2	&	2.9	$\pm$	0.1	&	8	:	11	:	17.2	$\pm$	0.01	&	-47	:	25	:	40	$\pm$	0.3	&	12.5	$\pm$	0.6	&	3.9	$\pm$	0.1	&	10.7	$\pm$	1.3	\\
G610-	4	&	8	:	11	:	16.6	$\pm$	0.01	&	-47	:	10	:	55	$\pm$	0.9	&	0.9	$\pm$	0.2	&	0.9	$\pm$	0.1	&	8	:	11	:	16.6	$\pm$	0.01	&	-47	:	10	:	55	$\pm$	0.9	&	11.8	$\pm$	2.1	&	3.4	$\pm$	0.2	&	12.1	$\pm$	4.4	\\
G610-	5	&	8	:	11	:	09.6	$\pm$	0.01	&	-46	:	57	:	49	$\pm$	0.6	&	2.2	$\pm$	0.3	&	2.3	$\pm$	0.1	&	8	:	11	:	09.6	$\pm$	0.01	&	-46	:	57	:	49	$\pm$	0.6	&	12.8	$\pm$	1.5	&	3.0	$\pm$	0.1	&	7.9	$\pm$	2.0	\\
G610-	6	&	8	:	11	:	03.8	$\pm$	0.01	&	-47	:	20	:	09	$\pm$	0.2	&	3.5	$\pm$	0.1	&	3.2	$\pm$	0.1	&	8	:	11	:	03.8	$\pm$	0.01	&	-47	:	20	:	09	$\pm$	0.2	&	12.3	$\pm$	0.5	&	3.4	$\pm$	0.0	&	10.6	$\pm$	0.8	\\
G610-	7	&	8	:	10	:	55.9	$\pm$	0.01	&	-47	:	20	:	13	$\pm$	0.5	&	1.3	$\pm$	0.1	&	1.2	$\pm$	0.1	&	8	:	10	:	55.9	$\pm$	0.01	&	-47	:	20	:	13	$\pm$	0.5	&	12.5	$\pm$	1.3	&	3.3	$\pm$	0.1	&	12.5	$\pm$	2.2	\\
G610-	8	&	8	:	10	:	55.7	$\pm$	0.01	&	-47	:	28	:	54	$\pm$	0.5	&	2.8	$\pm$	0.1	&	1.6	$\pm$	0.1	&	8	:	10	:	55.7	$\pm$	0.01	&	-47	:	28	:	54	$\pm$	0.5	&	14.1	$\pm$	1.1	&	4.6	$\pm$	0.2	&	13.5	$\pm$	2.8	\\
G610-	9	&	8	:	10	:	55.0	$\pm$	0.01	&	-47	:	28	:	58	$\pm$	0.2	&	8.8	$\pm$	0.1	&	3.7	$\pm$	0.1	&	8	:	10	:	55.0	$\pm$	0.01	&	-47	:	28	:	58	$\pm$	0.2	&	13.3	$\pm$	0.4	&	6.9	$\pm$	0.2	&	8.3	$\pm$	2.2	\\
G610-	10	&	8	:	10	:	31.2	$\pm$	0.01	&	-47	:	28	:	22	$\pm$	0.5	&	1.3	$\pm$	0.1	&	1.3	$\pm$	0.1	&	8	:	10	:	31.2	$\pm$	0.01	&	-47	:	28	:	22	$\pm$	0.5	&	12.0	$\pm$	1.1	&	3.2	$\pm$	0.1	&	9.9	$\pm$	1.9	\\
\hline
\end{tabular}
\tablefoot{$B_\mathrm{maj}$, $B_\mathrm{min}$ are the beam major and minor axes, and $B_\mathrm{pa}$ the beam position angle of the Gaussian fit by the pyBDSM routines.}
\end{sidewaystable*}

%-------------------------------------------------------------
%                                Table:  Sources w/ spix info
%-------------------------------------------------------------
\begin{table*}[h]
\caption{Sources found above 7$\sigma$ level at more than one frequency band, inside the 610-MHz field-of-view. Flux upper-limits were calculated as $7\times$r.m.s at the source position.} 
\label{tab:53sources}
\begin{tabular}{l@{~~~}r@{~~~}r@{~~~}r@{~~~}r@{~~~}r@{~~~}r@{~~~}r@{~~~}r}
\hline
\hline
ID & $RA_{\rm J2000}$ & $Dec_{\rm J2000}$ & $S_{\rm 150MHz}$  & $S_{\rm 325MHz}$  & $S_{\rm 610MHz}$ & $\alpha^{\rm 325MHz}_{\rm 150MHz}$ & $\alpha^{\rm 610MHz}_{\rm 325MHz}$ & $B_\mathrm{maj}$, $B_\mathrm{min}$, $B_\mathrm{pa}$ \\ 
&  (hms) & (dms) & (mJy) & (mJy) & (mJy) & & & ($''$,$''$,\degr)\\ 
\hline
I01& 08:07:01.9$\pm$0.01 & $-$47:28:52$\pm$0.8 & $<$8.2   & 3.1$\pm$0.28 & 4.5$\pm$0.45 &  $>-1.3$  & $+$0.6$\pm$0.50 & 22.3, 7.0, 11.3 \\ 
I02 & 08:07:17.8$\pm$0.01 & $-$47:37:07$\pm$0.1 & 206.2$\pm$2.0 & 87.5$\pm$0.51 & 35.4$\pm$0.53 & $-$1.1$\pm$0.03 & $-$1.4$\pm$0.06 & 23.0, 12.5, 9.5 \\ 
I03 & 08:07:26.2$\pm$0.01 & $-$47:21:26$\pm$0.2 & 21.4$\pm$2.0 & 11.5$\pm$0.27 & 5.5$\pm$0.28 & $-$0.8$\pm$0.28 & $-$1.2$\pm$0.20 & 23.7, 7.1, 7.5 \\
I04 & 08:07:29.0$\pm$0.01 & $-$47:19:51$\pm$0.6 & 13.8$\pm$2.0 & 7.1$\pm$0.25 & $<$1.1   & $-$0.9$\pm$0.44 & $<-2.9$ & 24.7, 10.2, 14.9 \\ 
I05 & 08:07:42.7$\pm$0.01 & $-$47:15:60$\pm$0.7 & $<$7.8  & 4.3$\pm$0.27 & 2.2$\pm$0.20 & $>-0.8$   & $-$1.1$\pm$0.41 & 24.6, 6.8, 7.8 \\ 
I06 & 08:07:51.7$\pm$0.01 & $-$47:13:49$\pm$0.1 & 83.1$\pm$1.9 & 33.9$\pm$0.28 & 11.9$\pm$0.21 & $-$1.2$\pm$0.07 & $-$1.7$\pm$0.07 & 23.2, 7.1, 8.6 \\
I07 & 08:08:01.9$\pm$0.01 & $-$47:37:06$\pm$1.2 & 62.2$\pm$1.4 & 8.2$\pm$0.29 & 3.7$\pm$0.36 & $-$2.6$\pm$0.13 & $-$1.3$\pm$0.38 & 37.3, 7.6, 6.6 \\ 
I08 & 08:08:08.5$\pm$0.01 & $-$47:32:35$\pm$0.1 & 1632.4$\pm$3.9 & 608.0$\pm$0.94 & 402.6$\pm$1.09 & $-$1.3$\pm$0.01 & $-$0.7$\pm$0.01 & 36.9, 11.9, 176.2 \\ 
I09 & 08:08:22.9$\pm$0.01 & $-$47:11:17$\pm$0.1 & 316.3$\pm$3.0 & 165.2$\pm$0.62 & 51.4$\pm$0.29 & $-$0.8$\pm$0.03 & $-$1.9$\pm$0.02 & 42.6, 22.9, 147.0 \\ 
I10 & 08:08:23.7$\pm$0.01 & $-$47:04:23$\pm$0.1 & 901.9$\pm$2.0 & 326.2$\pm$0.36 & 135.7$\pm$0.51 & $-$1.3$\pm$0.01 & $-$1.4$\pm$0.01 & 24.6, 7.2, 10.8 \\ 
I11 & 08:08:38.0$\pm$0.01 & $-$47:29:57$\pm$0.7 & $<$8.0   & 4.1$\pm$0.33 & 3.0$\pm$0.22 & $>-0.9$   & $-$0.5$\pm$0.40 & 22.5, 6.9, 9.1 \\ 
I12 & 08:08:38.3$\pm$0.01 & $-$47:18:09$\pm$0.9 & $<$7.7 & 2.1$\pm$0.25 & 0.9$\pm$0.16 & $>-1.7$   & $-$1.2$\pm$0.77 & 22.0, 6.3, 9.8 \\
I13 & 08:08:42.8$\pm$0.01 & $-$47:18:39$\pm$0.6 & $<$7.7 & 3.2$\pm$0.25 & 1.6$\pm$0.16 & $>-1.1$   & $-$1.0$\pm$0.45 & 22.2, 6.4, 9.1 \\ 
I14 & 08:08:44.5$\pm$0.01 & $-$47:42:33$\pm$0.5 & $<$8.5 & 6.2$\pm$0.30 & 4.8$\pm$0.28 & $>-0.4$   & $-$0.4$\pm$0.28 & 23.6, 6.9, 7.7 \\ 
I15 & 08:08:45.3$\pm$0.01 & $-$47:32:29$\pm$0.1 & 79.5$\pm$1.9 & 36.5$\pm$0.42 & 20.9$\pm$0.42 & $-$1.0$\pm$0.08 & $-$0.9$\pm$0.09 & 23.0, 12.7, 5.4 \\
I16 & 08:08:48.0$\pm$0.01 & $-$46:53:51$\pm$0.1 & 111.6$\pm$1.9 & 55.3$\pm$0.28 & 16.5$\pm$0.38 & $-$0.9$\pm$0.05 & $-$1.9$\pm$0.09 & 23.3, 6.6, 8.2 \\ 
I17 & 08:08:55.8$\pm$0.01 & $-$47:35:04$\pm$0.6 & $<$8.0 & 5.4$\pm$0.26 & 4.7$\pm$0.30 & $>-0.5$   & $-$0.2$\pm$0.29 & 24.9, 7.1, 8.3 \\ 
I18 & 08:09:09.6$\pm$0.02 & $-$47:29:19$\pm$1.2 & $<$7.8 & 2.0$\pm$0.24 & 0.8$\pm$0.19 & $>-1.8$   & $-$1.5$\pm$1.03 & 24.8, 6.4, 10.7 \\ 
I19 & 08:09:11.6$\pm$0.01 & $-$46:58:22$\pm$0.4 & 10.4$\pm$1.9 & 7.1$\pm$0.26 & 2.2$\pm$0.26 & $-$0.5$\pm$0.54 & $-$1.8$\pm$0.45 & 24.0, 7.3, 10.9 \\
I20 & 08:09:13.8$\pm$0.01 & $-$47:44:54$\pm$0.3 & 12.9$\pm$2.1 & 7.1$\pm$0.28 & 5.3$\pm$0.32 & $-$0.8$\pm$0.51 & $-$0.5$\pm$0.27 & 22.9, 6.6, 9.1 \\ 
I21 & 08:09:17.7$\pm$0.01 & $-$47:33:18$\pm$0.8 & $<$7.9 & 3.1$\pm$0.24 & 1.6$\pm$0.20 & $>-1.2$   & $-$1.1$\pm$0.54 & 24.1, 6.9, 7.8 \\
I22 & 08:09:25.6$\pm$0.01 & $-$47:41:30$\pm$0.4 & $<$8.3 & 6.6$\pm$0.25 & 4.9$\pm$0.27 & $>-0.3$   & $-$0.5$\pm$0.24 & 24.7, 6.9, 7.2 \\
I23 & 08:09:31.8$\pm$0.01 & $-$47:20:13$\pm$0.8 & $<$7.5  & 2.9$\pm$0.23 & 3.1$\pm$0.16 & $>-1.2$   & $+$0.1$\pm$0.35 & 23.3, 7.2, 6.7 \\ 
I24 & 08:09:32.6$\pm$0.01 & $-$47:05:13$\pm$0.4 & 8.0$\pm$1.9 & 5.6$\pm$0.23 & 1.7$\pm$0.19 & $-$0.5$\pm$0.72 & $-$1.9$\pm$0.43 & 23.5, 6.7, 7.2 \\ 
I25 & 08:09:43.0$\pm$0.01 & $-$47:14:26$\pm$0.4 & 22.5$\pm$1.7 & 4.9$\pm$0.23 & 2.1$\pm$0.15 & $-$2.0$\pm$0.26 & $-$1.4$\pm$0.31 & 22.7, 6.4, 8.1 \\ 
I26 & 09:09:43.6$\pm0.01$ & $-$47:19:38$\pm$0.1 & 331.1$\pm$2.7 & 140.8$\pm$0.32 & 67.7$\pm$0.36 & $-1.1 \pm 0.02$ & $-1.2 \pm 0.02$ & $\sim$60, 180, 10 \\
I27 & 08:09:43.6$\pm$0.01 & $-$47:33:28$\pm$0.6 & $<$7.7 & 3.0$\pm$0.24 & 1.4$\pm$0.18 & $>-1.2$   & $-$1.2$\pm$0.54 & 22.5, 6.4, 8.0 \\ 
I28 & 08:09:46.7$\pm$0.01 & $-$47:45:51$\pm$0.3 & 39.9$\pm$2.0 & 18.2$\pm$0.22 & 10.7$\pm$0.29 & $-$1.0$\pm$0.15 & $-$0.8$\pm$0.11 & 30.2, 11.1, 19.5 \\
I29 & 08:09:49.2$\pm$0.01 & $-$47:45:11$\pm$0.1 & 120.0$\pm$2.1 & 50.0$\pm$0.25 & 27.2$\pm$0.57 & $-$1.1$\pm$0.05 & $-$1.0$\pm$0.08 & 24.7, 10.9, 5.8 \\ 
I30 & 08:09:50.5$\pm$0.01 & $-$47:26:12$\pm$0.2 & 20.7$\pm$2.0 & 10.5$\pm$0.23 & 4.1$\pm$0.15 & $-$0.9$\pm$0.29 & $-$1.5$\pm$0.15 & 23.3, 6.6, 8.5 \\ 
I31 & 08:09:51.0$\pm$0.02 & $-$47:35:46$\pm$0.1 & 31.6$\pm$1.7 & 12.9$\pm$0.32 & 6.4$\pm$0.20 & $-$1.2$\pm$0.17 & $-$1.1$\pm$0.15 & 22.7, 9.2, 3.3 \\
I32 & 08:09:51.6$\pm$0.03 & $-$47:27:35$\pm$1.0 & 16.8$\pm$1.6 & 4.5$\pm$0.20 & $<$0.7 & $-$1.7$\pm$0.31 & $<-3.0$   & 25.7, 13.5, 10.0 \\ 
I33 & 08:09:54.6$\pm$0.01 & $-$47:30:32$\pm$0.2 & 46.4$\pm$1.9 & 17.3$\pm$0.21 & 6.9$\pm$0.21 & $-$1.3$\pm$0.13 & $-$1.5$\pm$0.12 & 25.0, 8.0, 4.5 \\
I34 & 08:09:55.0$\pm$0.01 & $-$47:37:21$\pm$0.6 & $<$7.9  & 5.8$\pm$0.22 & 2.3$\pm$0.19 &  $>-0.4$  & $-$1.5$\pm$0.33 & 25.9, 8.8, 6.1 \\ 
I35 & 08:09:55.7$\pm$0.02 & $-$47:31:26$\pm$0.1 & 20.6$\pm$1.9 & 9.9$\pm$0.30 & 3.3$\pm$0.15 & $-$1.0$\pm$0.29 & $-$1.7$\pm$0.20 & 24.9, 7.5, 5.7 \\
I36 & 08:10:00.1$\pm$0.01 & $-$47:05:06$\pm$0.1 & 90.6$\pm$1.8 & 44.4$\pm$0.32 & 18.7$\pm$0.23 & $-$0.9$\pm$0.06 & $-$1.4$\pm$0.05 & 23.8, 7.0, 7.8 \\ 
I37 & 08:10:02.6$\pm$0.01 & $-$47:11:31$\pm$0.7 & $<$7.2 & 3.0$\pm$0.22 & 1.4$\pm$0.16 & $>-1.1$   & $-$1.2$\pm$0.48 & 22.8, 7.4, 5.9 \\
I38 & 08:10:18.3$\pm$0.01 & $-$47:23:52$\pm$0.6 & $<$7.4 & 4.0$\pm$0.21 & 1.7$\pm$0.14 &  $>-0.8$  & $-$1.3$\pm$0.35 & 23.1, 7.8, 6.7 \\ 
I39 & 08:10:19.5$\pm$0.01 & $-$47:38:45$\pm$0.2 & $<$8.0 & 12.0$\pm$0.24 & 7.9$\pm$0.23 & $>0.5$   & $-$0.7$\pm$0.13 & 23.1, 7.2, 8.3 \\ 
I40 &  08:10:27.6$\pm$0.01 & $-$47:37:52$\pm$0.1 & 128.3$\pm$2.8 & 44.6$\pm$0.22 & 18.7$\pm$0.22 & $-$1.4$\pm$0.07 & $-$1.4$\pm$0.05 & 25.7, 10.0, 0.5 \\ 
I41 & 08:10:31.3$\pm$0.01 & $-$47:28:22$\pm$0.8 & $<$7.5 & 2.7$\pm$0.21 & 1.4$\pm$0.16 & $>-1.3$   & $-$1.0$\pm$0.48 & 23.7, 7.3, 6.1 \\ 
I42 & 08:10:37.4$\pm$0.01 & $-$47:18:01$\pm$0.7 & 12.8$\pm$1.8 & 5.0$\pm$0.19 & $<$0.7 & $-$1.2$\pm$0.42 & $<-3.1$  & 24.5, 11.8, 9.0 \\ 
I43 & 08:10:53.7$\pm$0.01 & $-$47:06:59$\pm$1.0 & 9.3$\pm$1.7 & 3.2$\pm$0.21 & $<$0.8 & $-$1.4$\pm$0.59 & $<-2.2$   & 26.9, 7.7, 6.9 \\ 
I44 & 08:10:55.2$\pm$0.01 & $-$47:28:58$\pm$0.1 & 47.8$\pm$1.8 & 25.4$\pm$0.20 & 12.8$\pm$0.15 & $-$0.8$\pm$0.11 & $-$1.1$\pm$0.05 & 24.9, 12.3, 12.9 \\ 
I45 & 08:10:56.0$\pm$0.01 & $-$47:20:12$\pm$0.6 & $<$7.2 & 2.9$\pm$0.22 & 1.4$\pm$0.16 &  $>-1.2$  & $-$1.2$\pm$0.49 & 22.3, 6.7, 6.9 \\ 
I46 & 08:11:03.8$\pm$0.01 & $-$47:20:10$\pm$0.3 & $<$7.2 & 7.8$\pm$0.22 & 5.3$\pm$0.17 & $>0.1$ & $-$0.6$\pm$0.15 & 22.9, 6.8, 7.78 \\ 
I47 & 08:11:09.6$\pm$0.01 & $-$46:57:48$\pm$0.2 & 39.1$\pm$1.8 & 13.0$\pm$0.26 & 2.4$\pm$0.34 & $-$1.4$\pm$0.15 & $-$2.7$\pm$0.51 & 23.2, 6.5, 9.0 \\
I48 & 08:11:16.6$\pm$0.01 & $-$47:10:55$\pm$0.6 & $<$7.0 & 4.0$\pm$0.23 & 1.0$\pm$0.20 & $>-0.7$   & $-$2.1$\pm$0.74 & 23.9, 6.8, 8.3 \\ 
I49 & 08:11:17.2$\pm$0.01 & $-$47:25:40$\pm$0.5 & $<$7.4 & 5.4$\pm$0.22 & 4.1$\pm$0.18 & $>-0.4$   & $-$0.4$\pm$0.22 & 24.1, 7.5, 4.1 \\  
I50 & 08:11:39.8$\pm$0.01 & $-$47:15:43$\pm$0.8 & $<$7.1 & 2.1$\pm$0.24 & 1.5$\pm$0.22 & $>-1.6$   & $-$0.6$\pm$0.69 & 20.9, 6.5, 7.3 \\ 
I51 & 08:11:46.1$\pm$0.01 & $-$47:26:14$\pm$0.6 & $<$7.4 & 4.1$\pm$0.24 & 2.4$\pm$0.24 & $>-0.8$   & $-$0.9$\pm$0.43 & 24.4, 7.0, 6.8 \\
I52 & 08:12:11.7$\pm$0.01 & $-$47:24:54$\pm$0.7 & 23.5$\pm$1.8 & 10.7$\pm$0.20 & $<$1.4 & $-$1.0$\pm$0.24 &  $<-3.2$  & 32.2, 12.4, 9.7 \\  
I53 & 08:12:17.0$\pm$0.01 & $-$47:23:36$\pm$0.4 & 48.5$\pm$1.8 & 14.8$\pm$0.22 & $<$1.4 & $-$1.5$\pm$0.12 & $<-3.7$    & 29.0, 11.6, 17.8 \\ 
\hline
\end{tabular}
\tablefoot{$B_\mathrm{maj}$, $B_\mathrm{min}$ are the beam major and minor axes at 325~MHz, and $B_\mathrm{pa}$ the beam position angle of the Gaussian fit by the pyBDSM routines. The source tagged here as I26 corresponds to the position of MOST\,0808--471 and is resolved in two and three components at 325 and 610~MHz respectively; see Table~\ref{tab:crossid} and also Sect. 4.2.}
\end{table*}
%-------------------------------------------------------------
%-------------------------------------------------------------
%                   Table: Cross-id of table-5 sources
%-------------------------------------------------------------
\begin{table*}[t!]
\caption{Cross identification of sources detected at more than one band.}
\label{tab:crossid}
\centering
\begin{tabular}{l r r r | l r r r | l r r r}  
\hline
\hline
ID & G325-- & G150-- & G610--& ID & G325-- & G150-- & G610--& ID & G325-- & G150-- & G610--\\
\hline
I01 & 190 & --- & 66 & I19 & 140 & 213 & 37 & I37 & 98 & --- & 16\\
I02 & 181 & 252 & 65 & I20 & 139 & 210 & 36 & I38 & 89 & --- & 14 \\
I03 & 179 & 250 & 64 & I21 & 134 & --- & 35 & I39 & 88 & --- & 13 \\
I04 & 178 & 248 & --- & I22 & 128 & --- & 34 & I40 & 84 & 169 & 11,12 \\
I05 & 173 & --- & 63 & I23 & 126 & --- & 33 & I41 & 81 & --- & 10 \\
I06 & 171 & 240 & 62 & I24 & 125 & 199 & 32 & I42 & 78 & 166 & --- \\
I07 & 169 & 239 & 61 & I25 & 120 & 195 & 30 & I43 & 72 & 162 & ---  \\
I08 & 167 & 236 & 53--60 & I26 & 114,115 & 192 & 26--28 & I44 & 71 & 161 & 8,9 \\
I09 & 161 & 229 & 48--50 & I27 & 119 & --- & 29 & I45 & 70 & --- & 7 \\
I10 & 162 & 230 & 47,51,52 & I28 & 111 & 185 & 24,25  & I46 & 67 & --- & 6 \\
I11 & 154 & --- & 45 & I29 & 110 & 186 & 23 & I47 & 64 & 157 & 5 \\
I12 & 155 & --- & 46 & I30 & 109 & 184 & 22 &  I48 & 61 & --- & 4 \\
I13 & 153 & --- & 44 & I31 & 108 & 182 & 21 &  I49 & 60 & --- & 3 \\
I14 & 152 & --- & 43 & I32 & 106 & 183 & --- &  I50 & 47 & --- & 2 \\
I15 & 151 & 221 & 41,42 & I33 & 104 & 181 & 20 &  I51 & 46 & --- & 1 \\
I16 & 150 & 220 & 40 & I34 & 103 & --- & 19 &  I52 & 36 & 148 & --- \\
I17 & 148 & --- & 39 & I35 & 102 & 180 & 18 & I53 & 33 & 137 & --- \\
I18 & 142 & --- & 38 & I36 & 99 & 179 & 17 &  & &  &  \\
\hline
\end{tabular}
\tablefoot{ID is the same as in Table~\ref{tab:53sources}; G325-- (G150--, G610--) is the name/number of the source detected at 325~MHz (150~MHz, 610~MHz) as in the Appendix Table .1.}
\end{table*}
%-------------------------------------------------------------

%-------------------------------------------------------------
%                   Table: Counterparts of sources w/ spix info
%-------------------------------------------------------------
\begin{table*}[t!]
\caption{Literature-searched sources at the position of GMRT sources with spectral index information.}
\label{tab:counterparts}
\centering
\begin{tabular}{l@{~~~}r@{~~~}l@{~~~}l@{~~~}l}
\hline
\hline
ID & $Dist$ & Nearest source name & Nearest source type & Comments \\
\hline
I03 &  $9.1''$ & 2MASS\,J08072621--4721354 & YSO candidate &\\
I08 & $2.8''$ & 2MASS\,J08080880--4732335 & YSO candidate  &\\
    & $13''$ & 2MASS\,J08080942--4732249 & YSO candidate &\\
I09 & $18.8''$ & 2MASS J08082235$-$4710594 & YSO & \textit{Fermi} excess \\
I23 & $1.9''$ & WR\,11 & WR star &  \textit{Fermi} excess \\
I26 & $7.3''$ & MOST 0808-471 & radio source & \textit{Fermi} excess \\
    &  $14.1''$ & 2MASS J08094219-4719526 & pre main-sequence star & \\ 
I36 & $7.2''$ & IRAS\,08084--4656 & IR point source & \textit{Fermi} excess \\
I49 & $10.3''$ & 2MASS\,J08111607--4711036 & YSO candidate  & \\
\hline
\end{tabular}
\tablefoot{ID: same as in Table~\ref{tab:53sources}; $Dist$: distance between the Simbad source and the detected source at the 325-MHz image; ``\textit{Fermi} excess'' means that the 325~MHz source lays within the \textit{Fermi} excess presented by \citet{2016MNRAS.457L..99P}.}
\end{table*}

\citet{2016PASA...33...17B} carried out a first search for sources at the 24-arcmin diameter ATCA field towards WR\,11, listing seven additional sources, including MOST\,0808--471. The GMRT images at 150, 325 and 610~MHz allow a deeper and more extended quest for counterparts. The FoVs corresponding to those bands are given in Table~\ref{Tab:1}.

We searched for radio continuum sources within each image using the Python Blob Detector and Source Finder (PyBDSF\footnote{http://www.astron.nl/citt/pybdsf/}). This tool decomposes a radio interferometric image into bi-dimensional Gaussians and provides with a list of sources and their parameters. In the process, the background r.m.s. is computed. When variations along the FoV are statistically significant, which is the case in our images, an r.m.s. map is built by estimating the r.m.s. value in few-pixels-size boxes; the detection threshold thus increases towards image edges. 
We have defined the threshold to separate source and noise pixels as 7r.m.s.\footnote{With this settings, the number of sources found on average were $\sim98$\% of those found for 5r.m.s.}. For checking the scope of the results, individual monitoring of random sources yielded no fake detections. Source sizes were not limited.

\begin{figure}
 \resizebox{\hsize}{!}{
 \includegraphics[width=\textwidth, angle=270]{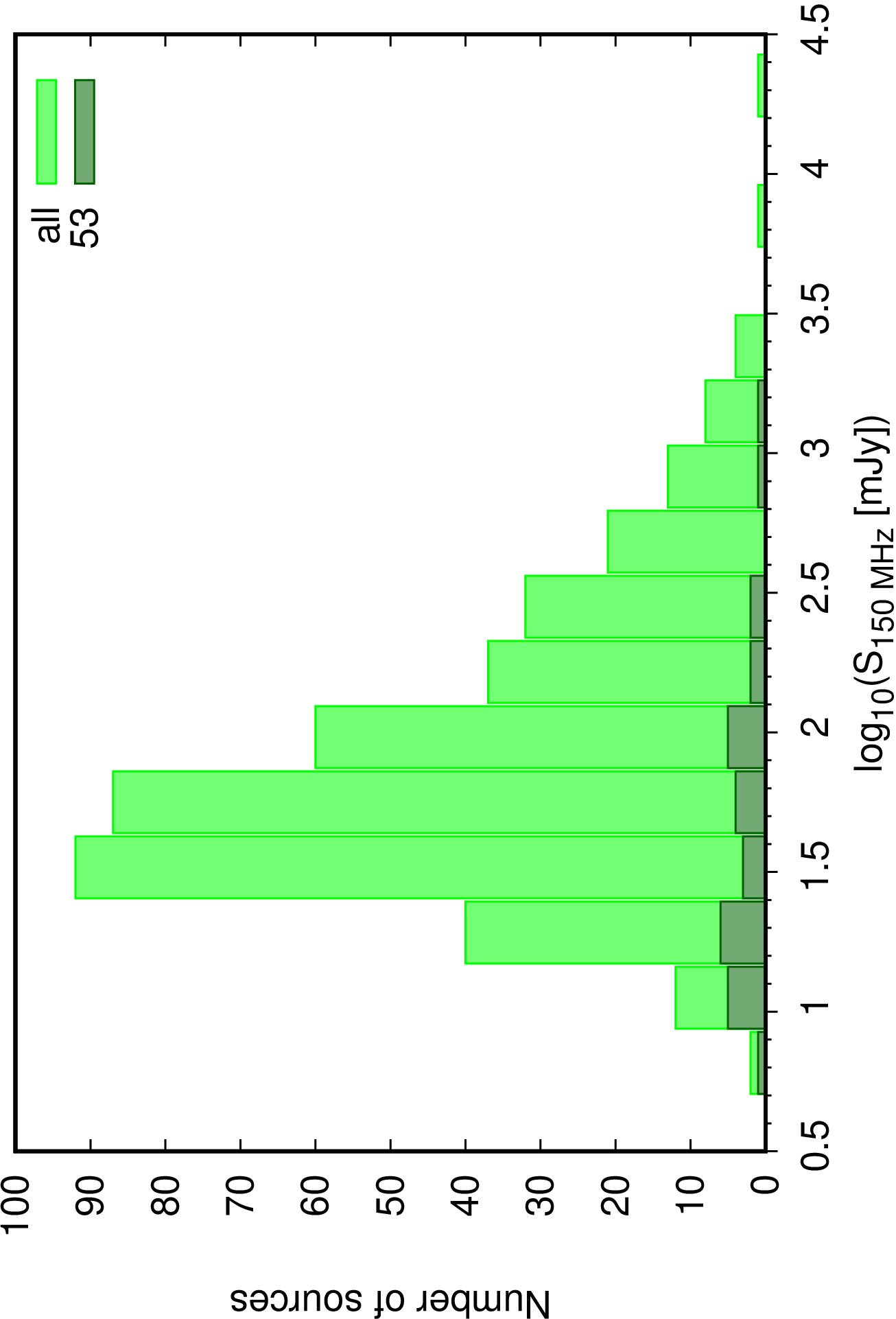}}\\
 \resizebox{\hsize}{!}{
 \includegraphics[width=\textwidth, angle=270]{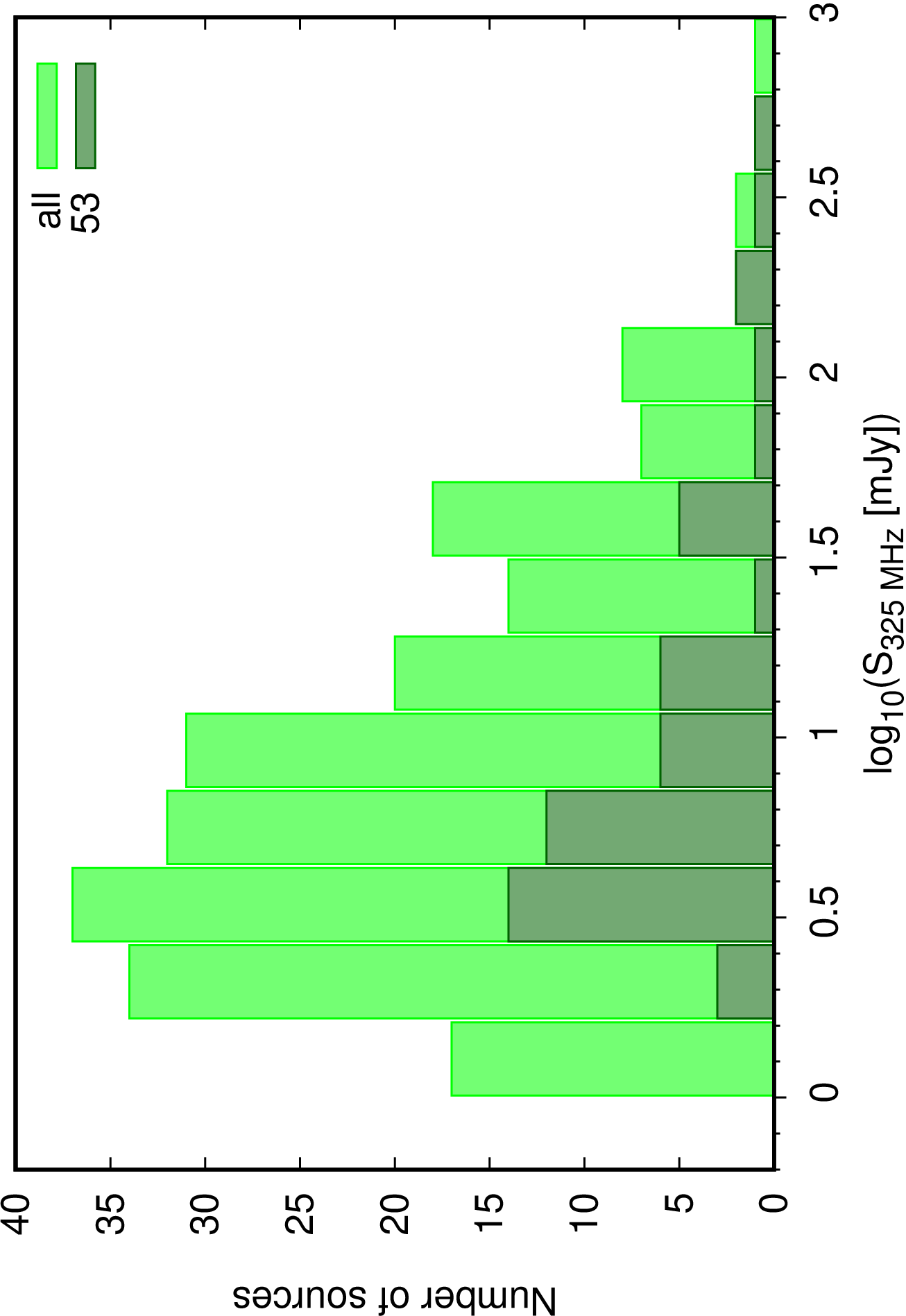}}
 \caption{Histograms of fluxes. Top: sources at 150~MHz. Bottom: sources at 325~MHz. Light green boxes are for the whole sample of sources detected at each wavelength, whereas in dark green the sample is restricted to the 53 sources described in Sect.~\ref{subsec:other_sources}. }
 \label{Fig:Histogram_flux}
\end{figure}

The PyBDSF run resulted in 410 sources at 150 MHz (tagged G150--1 to --410),  224 at 325~MHz (tagged G325--1 to --224) and  66 at 610~MHz (tagged G610--1 to --66). In Table~\ref{Tab:pyBSDFshort} we list the parameters of the first sources found at each band: ID (Column 1), coordinates and integrated flux of the fitted Gaussian functions (Cols. 2, 3 \& 4), the peak flux (Column 5), the coordinates of the emission maximum (Cols. 6 \& 7), and the major and minor axes and position angle of the fitted function (Cols. 8 to 10). The full list will be available as on-ine material. We also present in Fig.~\ref{Fig:Histogram_flux} histograms of the number of sources as a function of their flux at 150~MHz and 325~MHz.

%
%-------------------------------------------------------------
%                                   Figures 5: Fluxes histograms
%-------------------------------------------------------------
%-------------------------------------------------------------

\begin{figure}
 \resizebox{\hsize}{!}{
 \includegraphics[width=\hsize, angle=270]{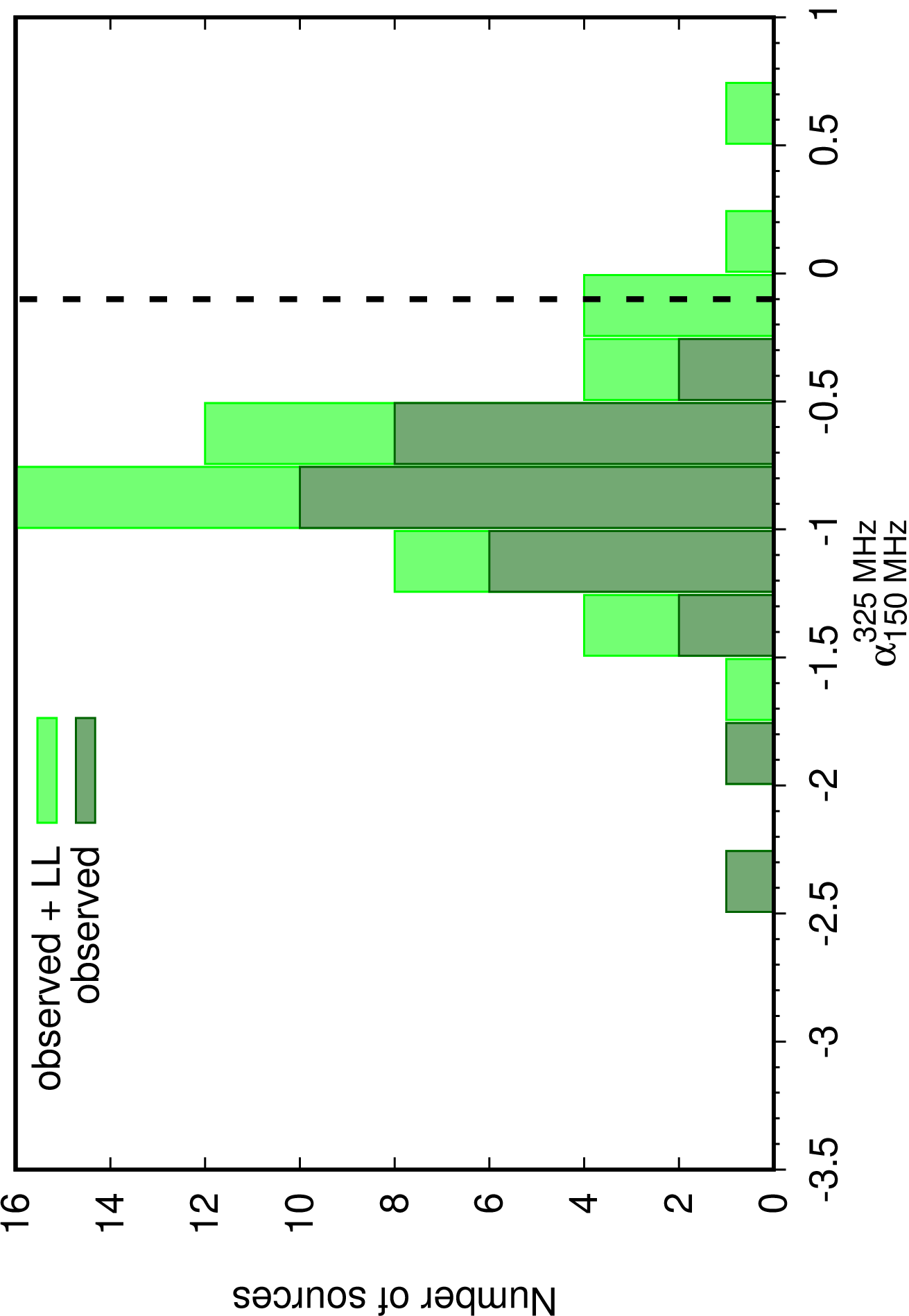}}\\ 
 \resizebox{\hsize}{!}{
 \includegraphics[width=\hsize, angle=270]{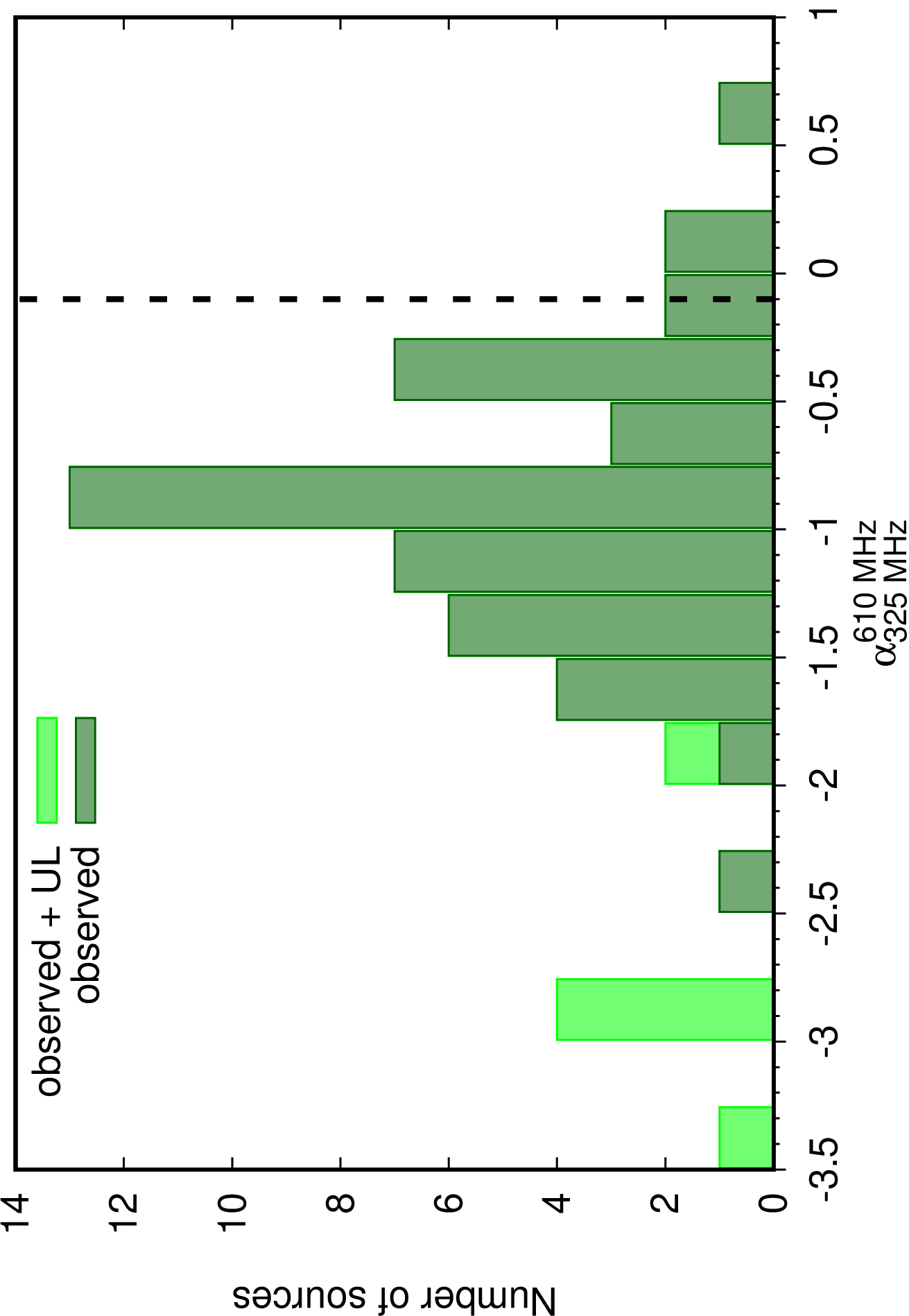}}
 \caption{Histograms of spectral indices using the values reported in Table~\ref{tab:53sources}. The dashed vertical line marks $\alpha = -0.1$, which is the lowest possible value for thermal emission. Top: between fluxes at 150 and 325~MHz; when only an upper limit to the flux at 150~MHz was available, we also estimated a lower limit (LL) to the spectral index. Bottom: between fluxes at 325 and 610~MHz; when only an upper limit to the flux at 610~MHz was available, we also estimated an upper limit (UL) to the spectral index.}
 \label{Fig:Histogram_spix}
\end{figure}

As a second step, we focused on the sources lying within a circular region of 43' in diameter, which corresponds to the smallest FoV of the three lower observing bands (610~MHz). The datasets produced by the PyBDSF tool allowed us to perform spectral index estimations. 
In this region we found 53 325-MHz sources with emission at 150~MHz and/or at 610~MHz above $7\sigma$ level. The sources are named as I01 to I53. 
The coordinates (Cols. 2 \& 3), fluxes (Cols. 4 to 6), spectral indices $\alpha$ (Cols. 7 \& 8) and sizes (fitted beam major and minor axes and position angle, Col. 9) are listed in Table~\ref{tab:53sources}, and are shown separately in the histograms of Fig.~\ref{Fig:Histogram_flux}. A distinction must be made regarding the spectral index values derived for these 53 sources. In the case in which the source fills the beam at both frequency bands, the spectral index corresponds to an average along the source regardless of its structure. The $\alpha$ values quoted for unresolved sources have to be taken with caution, since the pair of images to derive the spectral index were not corrected to the same beam. Such a detailed analysis will be presented elsewhere. In Fig.~\ref{Fig:Histogram_spix} we show histograms of the spectral indices obtained for these sources. 
In Table~\ref{tab:crossid} we list the cross-identifications among the sources found at more than one band (ID number, from I01 to I53) and the names given at the three different GMRT observed bands (G150--, G325-- and G610--). 

%-------------------------------------------------------------
%                                   Figure 6: Spixes Histograms
%-------------------------------------------------------------
%-------------------------------------------------------------

We carried out a Simbad search for counterparts over the region covered by the extension of the fitted source of Table 5 at the highest frequency (either 610 MHz when it was detected at that band, or 325 MHz for the same reason). Positive results are listed in Table~\ref{tab:counterparts}.  There we quote the source ID -same as in Table~\ref{tab:53sources}-, the distance to a Simbad possible counterpart, the name and type of this latter, and  whether the source lies within the \textit{Fermi} excess. In this way, in this last Table we indicate which sources with derived spectral indices were spatially coincident with the \textit{Fermi} excess reported by \citet{2016MNRAS.457L..99P}; there are 21 of such, over the area portrayed by \citet{2016MNRAS.457L..99P} and \citet{2016PASA...33...17B}.

%--------------------------------------------------Sect. 6-----------------
\section{Discussion}
\label{Sec:discussion}

\subsection{The WR\,11 system}

According to \citet{2017ApJ...847...40R}, the reported $\gamma$-ray emission by \citet{2016MNRAS.457L..99P} can only be achieved by WR\,11 in a hadronic model, as the predicted leptonic $\gamma$-ray emission is far too weak. The authors show that the $\gamma$-ray spectrum can be fitted with a p-p component for certain combinations of an energy-dependent diffusion coefficient in the WCR, a proton injection ratio 
$\eta_\mathrm{p} \sim 10^{-3}$, and considering a mass-loss rate of the WR star of $\dot{M}_{WR} = 3 \times 10^{-5}$~M$_\odot$~yr$^{-1}$. However, the authors note that a smaller value of $\dot{M}_{WR} = 8 \times 10^{-6}$~M$_\odot$~yr$^{-1}$ leads to an unrealistically high energetics requirement. It is thus relevant to explore deeper the issue of the actual mass loss rate.

Our set of measurements consistently point to a thermal origin for the radio emission from WR\,11 from 0.3 to 230~GHz, refuting previous suggestions of the presence of a NT component at low frequencies \citep{1999ApJ...518..890C}. The SED shown in Fig.~\ref{Fig:SED_WR11} does not present any hint for a NT emission component contributing to the measured radio flux, which can thus be interpreted as optically thick thermal emission from stellar winds as described, for instance, by \citet{1975MNRAS.170...41W}. First of all, one should keep in mind that the WR\,11 system is made up of two stars, both of them likely to contribute to the thermal free-free emission. Let us estimate the potential contribution of the O-star to the measured flux densities, using the relations published by \citet{1975MNRAS.170...41W}. According to \citet{2017MNRAS.468.2655L}, the mass of the O-star companion should be about 28 solar masses. For an O7.5 spectral type, according to the calibration of O-type star parameters published by \citet{2005A&A...436.1049M}, this points to a giant luminosity class. We note, however, that \citet{2017MNRAS.468.2655L} suggest that the best-matching IR spectrum is that of an O6.5I star, but such a spectral classification would be at odd with the mass of about 28 solar masses which should be quite robust. In addition, let us caution that the O-type spectrum analyzed by \citet{2017MNRAS.468.2655L} is very likely contaminated by some emission from the colliding-wind region. This contamination casts some doubt on any luminosity class determination based only on the IR spectrum. We will therefore assume the O7.5III classification for the O-star. Adopting the mass loss rate and terminal velocity proposed by \citet{2012A&A...537A..37M} for that spectral classification, one obtains a flux density lower than 0.1~mJy at 1.4~GHz. In addition, let us note that the free-free emission from the O-star wind should be significantly reduced considering that a significant part of that wind is smashed by the collision with the WR wind. This is supposed to lead to a further reduction of the free-free emission from that wind with respect to our estimate.  Noting that the measured flux density at 1.4~GHz is 10~mJy, one can conclude that the radio measurements are clearly dominated by the free-free emission from the WC8 wind.

We thus estimated the mass loss rate of the WC wind using the following equation (adapted from \citealt{1975MNRAS.170...41W}),
$${\dot M} = 4\,\pi\mu\,m_H\,v_\infty\,\bigg[\frac{2}{\pi\,K_\nu\,\gamma}\bigg]^{1/2}\,\bigg[\frac{S_\nu}{f_{cl}^{2/3}}\,\frac{d^2}{1.34\,2\pi\,B_\nu}\bigg]^{3/4}.$$
\noindent The expression for $K_\nu$ can be found in \citet{1975MNRAS.170...41W}, and $B_\nu$ stands for Planck's function which can be considered in the Rayleigh-Jeans limit at radio frequencies. Assuming a WC wind made only of He, we adopted a mean molecular weight ($\mu$) value of 4.0, along with an r.m.s. ionic charge ($\gamma$) of 1.0 \citep[for details, see][]{1995ApJ...450..289L}. The clumping factor ($f_{cl}$) was assumed to be equal to 4.0, in agreement with the predictions for outer parts of massive star winds made by \citet{2002A&A...381.1015R}, where the measured free-free radio emission is coming from\footnote{As discussed notably by \citet{2008A&ARv..16..209P}, mass loss rates corrected for the effect of wind clumping are reduced by a factor $\sqrt{f_{cl}}$. As the flux density is proportional to the mass loss rate to the 4/3 power \citep{1975MNRAS.170...41W}, this translates into a correction of flux densities by a factor $f_{cl}^{2/3}$. For $f_{cl} = 4$, we achieve the expected reduction of the mass loss rate by a factor $\sqrt{2}$.}. $T$ is the electron temperature set to be 50$\%$ of the effective temperature \citep{1990ASPC....7..230D}. The latter temperature for a WC8 star is expected to be about 70000~K \citep[see ][]{2007ARA&A..45..177C}. We adopted a terminal velocity ($v_\infty$) of 2000~km~s$^{-1}$. Following this approach, we obtain a mass loss rate for the WC star of the order of $2.4 \times 10^{-5}$~M$_\odot$~yr$^{-1}$. This value is very close to the value adopted by \citet{2017ApJ...847...40R} for their simulations (see above in this section). Such a value seems thus high enough to provide the required kinetic power to significantly feed NT emission processes.

\begin{figure}
    \resizebox{\hsize}{!}{
	\includegraphics[width=\hsize]{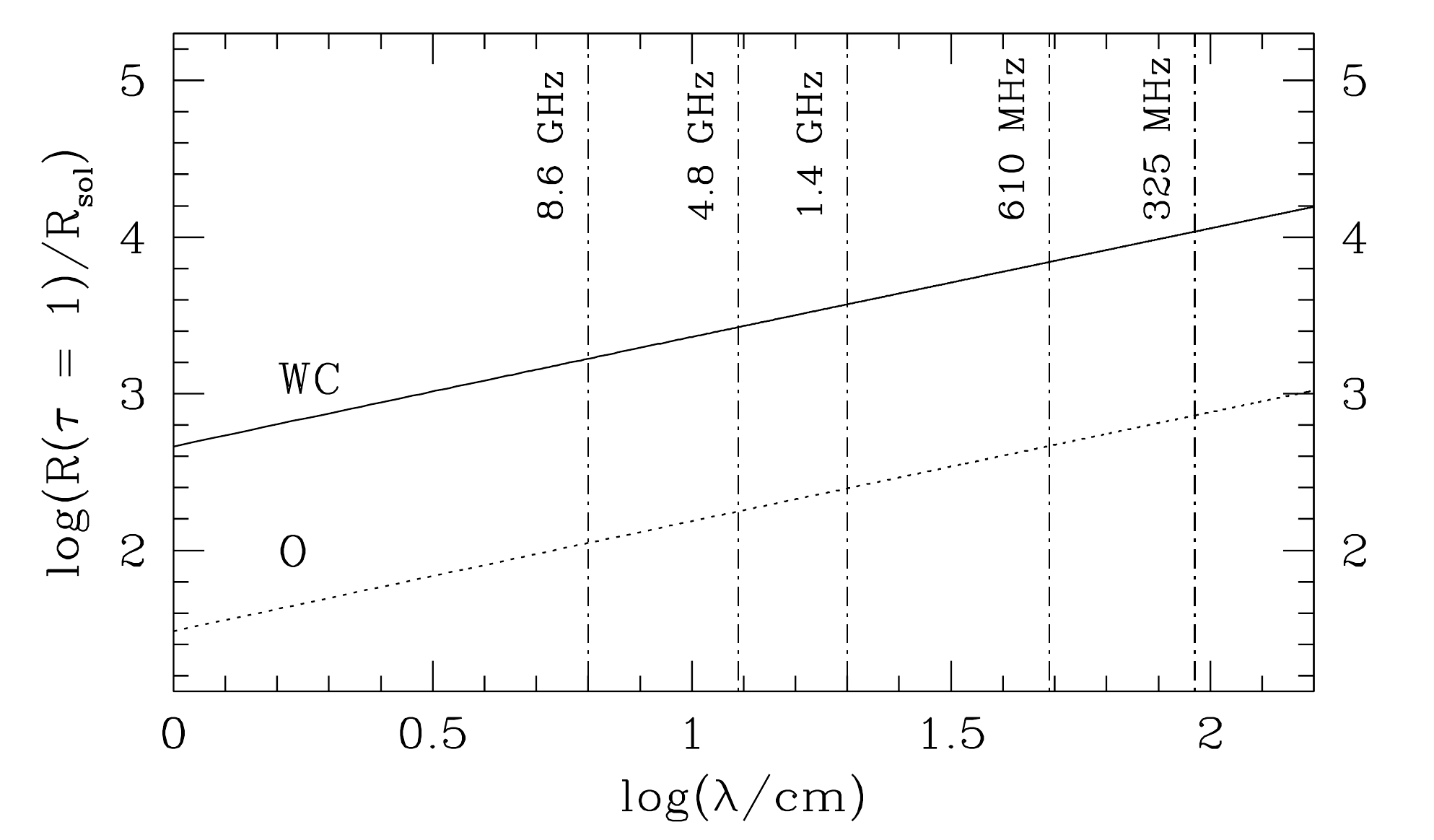}}
    \caption{Radio photosphere of the stellar winds in WR\,11 as a function of radio wavelength. Vertical dashed lines emphasize some characteristic frequencies specified in the plot. The thick and the dotted lines represent the photosphere sizes of the WC and the O components, respectively. 
    }
    \label{Fig:radiophotwr11}
   \end{figure}

In the framework of the potential contribution of WR~11 to the $\gamma$-ray emission reported on by \citet{2016MNRAS.457L..99P}, the lack on NT radio emission deserves some comments. In quite short period binary systems, free-free absorption (FFA) is expected to severely reduce any putative synchrotron radio emission produced in the WCR. In order to achieve a view of the capability of the stellar winds in the system to absorb significantly radio photons, we calculated the radius of the so-called radio photosphere for an FFA radial optical depth equal to one in a wide range of photon wavelength. This calculation is performed using equations given by \citet{1975MNRAS.170...41W}, adopting the same approach as De Becker et al. (2019, submitted) 
for the massive binary WR~133. The curves were computed for both stellar winds adopting the parameters specified above in this section, and are shown in Fig.~\ref{Fig:radiophotwr11}. First, it is clear that FFA in the system is dominated by the WC wind. Second, the extension of the high optical depth zone is much larger than the typical dimension of the full binary system. According to \citet{2017MNRAS.468.2655L}, the semimajor axis of the orbit is about 3.5 milli-arseconds, which translates to a linear semimajor axis of about 250~R$_\odot$ at a distance of about 340~pc (in agreement with the results published by \citealt{2007MNRAS.380.1276N}). This is much smaller than the expected size of the radio photosphere at our selected wavelength, which is much larger than a few 1000~R$_\odot$ at all frequencies below 300~GHz (Fig.~\ref{Fig:radiophotwr11}). The putative synchrotron emitting region in the system is thus completely and deeply embedded in the high FFA opacity region. Any synchrotron emission originating from the WCR would therefore be completely absorbed by the opaque WC wind. As a result, the lack of evidence for NT emission in the radio domain cannot be interpreted as an evidence for the absence of any NT process at work in the system. This issue has already been discussed by \citet{2017A&A...600A..47D}, as for instance in the case of the CWB $\eta$-Car both NT X rays and $\gamma$ rays have been detected but not NT radio emission.

%-------------------------------------------------------------
%                                 Figure 7: Radio photospheres
%-------------------------------------------------------------

\subsection{MOST\,0808--471}
\label{disc-most}

The SED presented in Fig.~\ref{Fig:SED_MOST0808} was first fitted by a power-law, taking into account all the flux densities available for this source. The derived spectral index is $\alpha = -0.97 \pm 0.09$, which is consistent with NT synchrotron emission produced by a population of relativistic electrons with a power-law energy distribution with index $p = 1 - 2\,\alpha \approx -2.9$ (for a distribution defined as $N(E) \propto E^p$, where $E$ is the electron energy). Such a steep synchrotron radio spectrum may potentially be explained by a population of relativistic electrons accelerated by shocks (through the DSA mechanism) characterized by a compression ratio lower than the expected value for strong, adiabatic shocks. In the latter case, the compression ratio is expected to be $\chi = 4$. The linear DSA process predicts thus an electron index $p = -\frac{\chi +2}{\chi -1} = -2$. However, any compression ratio lower than 4 will lead to an electron index lower than $-2$, and consequently a steeper synchrotron spectrum. Such a situation happens for instance in young supernova remnants, where the high efficiency of particle acceleration leads to a significant feedback of the particle acceleration on the shock properties. This results in a deceleration of the upstream flow close to the shock discontinuity, leading to a reduction of the velocity jump, and accordingly to a reduction of the effective compression ratio \citep[see e.g.][for a discussion of supernova remnants]{2012SSRv..166..231R}. This effect -- referred to as shock-modification -- may thus in principle explain a significant deviation with respect to the canonical index resulting from DSA-generated relativistic electron populations. If this interpretation applies to MOST\,0808--471, this would indicate a quite high particle acceleration efficiency. Otherwise, shock-modification would not be significant.

However, the data plotted in the SED presented in Fig.~\ref{Fig:SED_MOST0808} suggest a slightly flatter spectrum below 4.8\,GHz, followed by a steeper spectrum at higher frequencies. The determination of the spectral index below 4.8\,GHz yields $\alpha = -0.91 \pm 0.06$ (translating into $p \approx -2.8$), which is still quite steep. The unusual values of the spectral index, along with the apparent steepening at higher radio frequencies may point also to an inhomogeneous distribution of the magnetic energy density across the emitting region. One knows that the typical synchrotron photon frequency is proportional to the magnetic field strength and to the square of the emitting electron energy \citep[see e.g.][]{1979rpa..book.....R}. For a given population of relativistic electrons distributed across the source, different parts of the emitting region characterized by different magnetic field strengths may affect the measured
5
 synchrotron emission that is averaged over the full emitting region. A larger emitting volume with a lower magnetic field will lead to a stronger contribution at lower frequencies, whilst a smaller volume with a stronger magnetic field will drive the emission from a smaller amount of relativistic electrons at higher frequencies, thus less contributing to the spectrum.

Alternatively, one should also consider the gradual cooling of relativistic electrons, with the more energetic electrons radiating at a higher rate than less energetic ones. This may lead to a gradual steepening of effective electron distribution whose signature is measured through the synchrotron spectrum. Depending on the source properties, the cooling may be only due to synchrotron radiation, or may also be influenced by inverse Compton scattering in the Thomson regime. However, in the absence of clear indication on the nature of MOST\,0808--471, one can only speculate on the interpretation of the measured radio spectrum.

%-------------------------------------------------------------
%                                   Figures 8: MOST lobe fluxes
%-------------------------------------------------------------
\begin{figure}
 \resizebox{\hsize}{!}{
 \includegraphics[width=\hsize, angle=0]{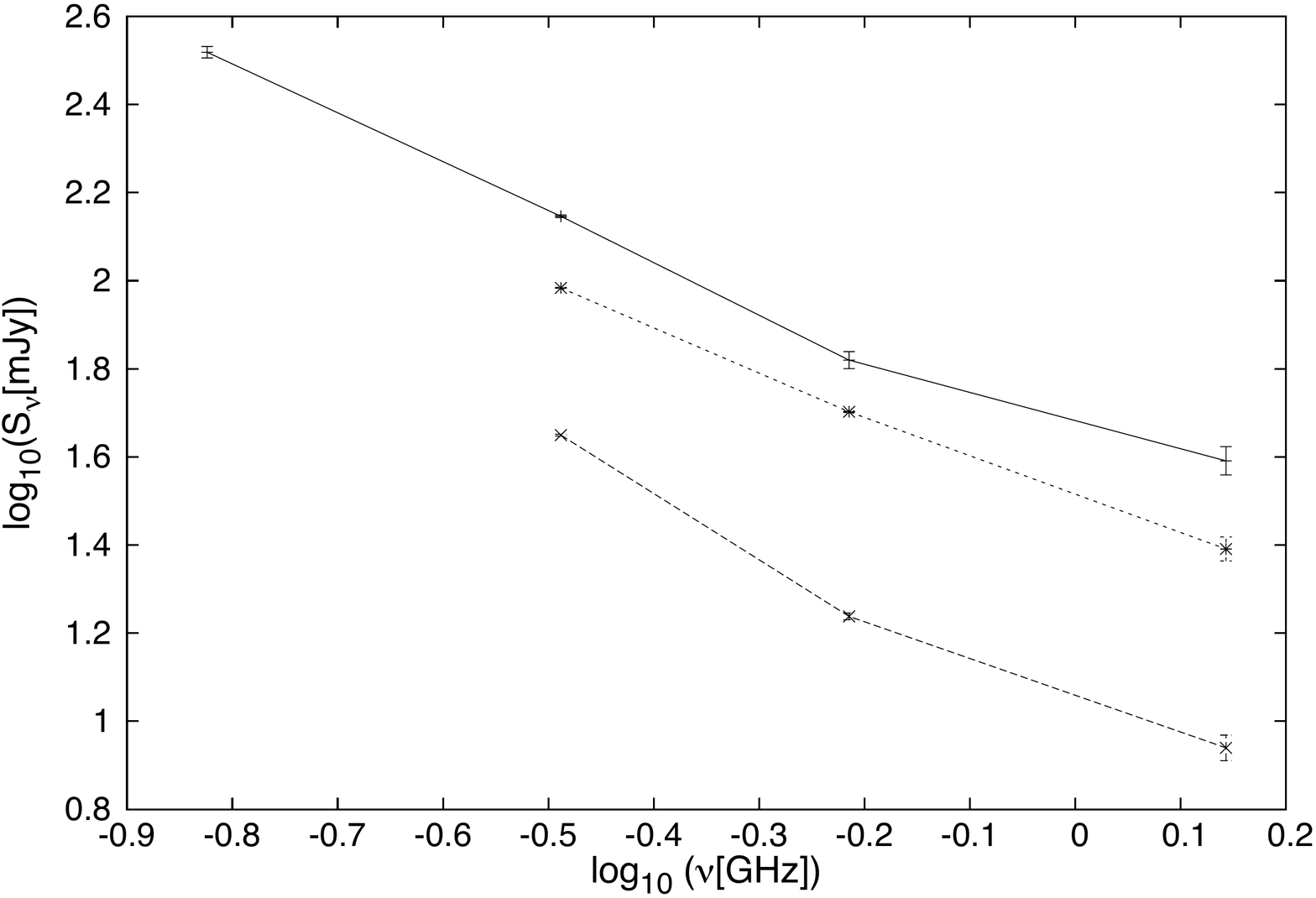}}
 \caption{GMRT fluxes of the MOST\,0808--471 counterpart components. Dashed line: left component; dotted line: right component; solid line: total flux.}
 \label{Fig:most-lobe-fluxes}
\end{figure}
%-------------------------------------------------------------

The images at the position of MOST\,0808--471 obtained from the GMRT observations presented here are the most sensitive with the highest angular resolution at radio-cm frequencies.
In Figs.~\ref{Fig:s5-6a-layout} \& ~\ref{Fig:s5-6b-layout} GMRT counterparts can be identified. At $\nu \ge 325$~MHz, MOST\,0808--471 appears resolved in more than one components, (all of) which, in principle, may or may not be related to the same physical object: a double-lobe single object or a set of two line-of-sight objects are possible. In Table~\ref{Tab:most-table} we discriminate the fluxes computed for left and right counterparts (named left and right lobes for simplicity), see also Fig.~\ref{Fig:most-lobe-fluxes}. The spectral indices are $\overline{\alpha} \approx -1.2$ and  $\overline{\alpha} \approx -0.9$ for the left and the right lobes.
Provided the two 'lobes' are physically related, a straightforward hypothesis is that MOST\,0808--471 consists of a NT bipolar jet source. 
Such an object is expected in various well-known scenarios. For instance, \citet{2016ApJ...818...27R} described the synchrotron emission from YSO jets \citep[see also the review by][]{2018A&ARv..26....3A}, and complemented the study with theoretical models that predict that these objects can produce $\gamma$-ray emission as well, involving bipolar jets \citep[e.g.,][]{2007A&A...476.1289A}. Microquasars were also widely studied as $\gamma$-ray emitters with two-lobed radio counterparts \citep{2003A&A...410L...1R}. If the two lobes related to MOST\,0808--471 belong to a single object at a distance of 360~pc, their projected separation is $\sim 5000$~AU. Another possibility is that we are dealing with an extragalactic source such as a quasar. These objects are well known to be NT emitters from radio to $\gamma$ rays \citep[see, for instance,][]{2016A&ARv..24...10T,2015ApJ...810...14A,2017AIPC.1792b0008R}.

At the present stage, the information compiled on MOST\,0808--471 turns it a promising candidate to contribute to the GeV excess found by \citet{2016MNRAS.457L..99P}. However, given the generally poor angular resolution of present $\gamma$-ray instruments, carrying multi-wavelength observations is of paramount importance and should always be considered before jumping into conclusions about the nature of \textit{Fermi} sources, in particular in the case of the field of WR\,11 that is rich in non-thermal sources.

\subsection{Radio detected sources}

The areas surveyed to produce the catalog of sources listed as material accompanying this article 
correspond to the FoV circles (defined as the HPBW) of the GMRT at the observed bands, centred at WR~11 with diameters of 3$^\circ$ at 150~MHz, 1.5$^\circ$ at 325~MHz, and 0.75$^\circ$ at 610~MHz. The detection threshold is given by the image r.m.s. (see Table~\ref{Tab:1}).

The number of sources detected at 150, 325 and 610~MHz resulted in 410, 224 and 66 respectively, yielding densities of 14, 39 and 41 sources --above 7r.m.s.-- per square degree. The discrete sources at each band corresponded to  78\%, 82\% and 69\% of the sources fitted.

Of the 53 sources detected at least in two bands, 24 are detected at the three bands; 6 at only the 150 and 325~MHz bands and 24 at only the 325 and 610~MHz bands. Of this last group, four of them presented spectral index consistent with -or dominated by- thermal emission, 
whereas the rest showed spectral indices below $-0.1$ and are, therefore, most likely dominated by NT emission. In addition, 19 of these  53 sources are projected onto the \textit{Fermi} excess region. The search for counterparts of 19 sources resulted in correlations with stellar sources, like young stellar objects (candidates), stars, and pre main-sequence stars.

From a theoretical point of view, at low frequencies {\sl (i)} the NT emission (if existent) should be dominant, so one could expect that NT spectral indices should be very frequent, whereas spectral indices obtained at larger frequencies could be less negative due to ``contamination'' with thermal emission that becomes more relevant. Nonetheless, {\sl (ii)} absorption/suppression effects might be important and these affect more at low frequencies; the consequent decrease of the low-frequency emission implies that the spectral index obtained from the low frequency range might be more positive than the one obtained at larger frequencies. In fact, that tendency can be appreciated in Fig.~\ref{Fig:Histogram_spix}.

Fig.~\ref{Fig:Histogram_spix} (bottom) shows that almost all sources present a NT spectral index when estimated from the 325~MHz and 610~MHz fluxes, suggesting that any suppression/absorption effects that could be at work at those frequencies are not efficient/relevant. In contrast, Fig.~\ref{Fig:Histogram_spix} (top) shows that spectral indices estimated from the 325~MHz fluxes and the 150~MHz flux upper-limits also present a small number of sources with positive spectral indices. We interpret this as evidence that, at least for some sources, suppression/absorption effects could be more relevant at the lowest frequencies of this study.
For the present study, the images were built to emphasize point sources, thus some part of the diffuse (possibly thermal) emission is not shown. 
The fact that the majority of the sources in Table~\ref{tab:53sources} have NT spectral indices is probably due to an observational bias, as NT sources are stronger than thermal ones at low frequencies. Thus, the NT contribution prevails regardless of whether both contributions, thermal and NT, are present. Moreover, at lower frequencies the r.m.s. increases so less weak sources are revealed. The content of Table~\ref{tab:53sources} results therefore from a combination of decreasing sensitivity threshold with frequency -and also angular resolution-, and a weakening of thermal emission at lower frequencies with respect to the NT one.

%--------------------------------------------------Sect. 6-----------------
\section{Conclusions}
\label{Sec:conclusions}

We studied the massive binary system WR\,11 and its surroundings by means of dedicated radio interferometric observations from 150~MHz to 1.4~GHz, plus archive data up to 230~GHz. Our data set, in particular the new GMRT data, constitute to date the most sensitive radio survey of this region, with unprecedented angular resolution.

Our analysis allowed us to draw the following inferences:
\begin{enumerate}
\item[$\bullet$] From our GMRT observations we identified more than 400 radio emitters in the vicinity of WR~11, among which a significant fraction is detected in more than one band. A large fraction of them present a non-thermal spectrum, in agreement with the expectations at such low frequencies. We also show evidence suggesting that in many sources absorption/suppression processes significantly shape the spectral energy distribution at frequencies below 325~MHz.

\item[$\bullet$] The radio spectrum of WR~11 confirms the predominance of thermal emission from 150~MHz to 230~GHz. The results presented here constitute the only set of measurements leading to a spectral index determination down to 150~MHz 

for a colliding wind binary to date. In this short-period binary system, the WCR is buried in the radio photospheres of the binary components, and non-thermal emission from it is expected to be suppressed by strong free-free absorption from the individual thermal winds. These results strongly suggest that the interpretation by \citet{1999ApJ...518..890C} of a non-thermal component in the radio spectrum of WR\,11 is misleading.

\item[$\bullet$] The measured fluxes allowed us to derive a stellar mass-loss rate that is enough, according to the model by \citet{2017ApJ...847...40R}, to explain a putative contribution of WR\,11 to the \textit{Fermi} excess. 

\item[$\bullet$] Our analysis allowed partial characterization of close-by source(s) at the position of MOST\,0808-471 that presents intense, non-thermal radio emission, confirming its capability to participate in non-thermal processes. As a result, MOST\,0808-471 deserves to be considered as a potential contributor to the $\gamma$-ray source identified by \citet{2016MNRAS.457L..99P}. A detailed multi-wavelength study however is needed to investigate its nature and potential capability to radiate at GeV energies.
\end{enumerate}

\begin{acknowledgements}
We thank the staff of the GMRT that made these observations possible. GMRT is run by the National Centre for Radio Astrophysics of the Tata Institute of Fundamental Research. P. B. thanks Divya Overoi and Huib Intema for fruitful discussions on data reduction during her stay at the NCRA facility and Niruj Moham Ramanujam and Marcelo E. Colazo for help related with the pyBDSF tool handling. S. del P. thanks the Stack Exchange community for the useful information available. This research has made use of the SIMBAD database, operated at CDS, Strasbourg, France and of NASA's Astrophysics Data System. 
\end{acknowledgements}

 \bibliographystyle{aa} % style aa.bst
 \bibliography{wr11} % your references Yourfile.bib

\begin{thebibliography}{51}
\expandafter\ifx\csname natexlab\endcsname\relax\def\natexlab#1{#1}\fi

\bibitem[{{Ackermann} {et~al.}(2015){Ackermann}, {Ajello}, {Atwood}, {Baldini},
  {Ballet}, {Barbiellini}, {Bastieri}, {Becerra Gonzalez}, {Bellazzini},
  {Bissaldi}, {Bland ford}, {Bloom}, {Bonino}, {Bottacini}, {Brandt},
  {Bregeon}, {Britto}, {Bruel}, {Buehler}, {Buson}, {Caliandro}, {Cameron},
  {Caragiulo}, {Caraveo}, {Carpenter}, {Casandjian}, {Cavazzuti}, {Cecchi},
  {Charles}, {Chekhtman}, {Cheung}, {Chiang}, {Chiaro}, {Ciprini}, {Claus},
  {Cohen-Tanugi}, {Cominsky}, {Conrad}, {Cutini}, {D'Abrusco}, {D'Ammando}, {de
  Angelis}, {Desiante}, {Digel}, {Di Venere}, {Drell}, {Favuzzi}, {Fegan},
  {Ferrara}, {Finke}, {Focke}, {Franckowiak}, {Fuhrmann}, {Fukazawa},
  {Furniss}, {Fusco}, {Gargano}, {Gasparrini}, {Giglietto}, {Giommi},
  {Giordano}, {Giroletti}, {Glanzman}, {Godfrey}, {Grenier}, {Grove},
  {Guiriec}, {Hewitt}, {Hill}, {Horan}, {Itoh}, {J{\'o}hannesson}, {Johnson},
  {Johnson}, {Kataoka}, {Kawano}, {Krauss}, {Kuss}, {La Mura}, {Larsson},
  {Latronico}, {Leto}, {Li}, {Li}, {Longo}, {Loparco}, {Lott}, {Lovellette},
  {Lubrano}, {Madejski}, {Mayer}, {Mazziotta}, {McEnery}, {Michelson},
  {Mizuno}, {Moiseev}, {Monzani}, {Morselli}, {Moskalenko}, {Murgia}, {Nuss},
  {Ohno}, {Ohsugi}, {Ojha}, {Omodei}, {Orienti}, {Orland o}, {Paggi},
  {Paneque}, {Perkins}, {Pesce-Rollins}, {Piron}, {Pivato}, {Porter},
  {Rain{\`o}}, {Rando}, {Razzano}, {Razzaque}, {Reimer}, {Reimer}, {Romani},
  {Salvetti}, {Schaal}, {Schinzel}, {Schulz}, {Sgr{\`o}}, {Siskind},
  {Sokolovsky}, {Spada}, {Spandre}, {Spinelli}, {Stawarz}, {Suson},
  {Takahashi}, {Takahashi}, {Tanaka}, {Thayer}, {Thayer}, {Tibaldo}, {Torres},
  {Torresi}, {Tosti}, {Troja}, {Uchiyama}, {Vianello}, {Winer}, {Wood}, \&
  {Zimmer}}]{2015ApJ...810...14A}
{Ackermann}, M., {Ajello}, M., {Atwood}, W.~B., {et~al.} 2015, \apj, 810, 14

\bibitem[{{Anglada} {et~al.}(2018){Anglada}, {Rodr{\'\i}guez}, \& {Carrasco-
  Gonz{\'a}lez}}]{2018A&ARv..26....3A}
{Anglada}, G., {Rodr{\'\i}guez}, L.~F., \& {Carrasco- Gonz{\'a}lez}, C. 2018,
  Astronomy and Astrophysics Review, 26, 3

\bibitem[{{Araudo} {et~al.}(2007){Araudo}, {Romero}, {Bosch-Ramon}, \&
  {Paredes}}]{2007A&A...476.1289A}
{Araudo}, A.~T., {Romero}, G.~E., {Bosch-Ramon}, V., \& {Paredes}, J.~M. 2007,
  \aap, 476, 1289

\bibitem[{{Benaglia}(2016)}]{2016PASA...33...17B}
{Benaglia}, P. 2016, Publications of the Astronomical Society of Australia, 33,
  e017

\bibitem[{{Chapman} {et~al.}(1999){Chapman}, {Leitherer}, {Koribalski},
  {Bouter}, \& {Storey}}]{1999ApJ...518..890C}
{Chapman}, J.~M., {Leitherer}, C., {Koribalski}, B., {Bouter}, R., \& {Storey},
  M. 1999, \apj, 518, 890

\bibitem[{{Crowther}(2007)}]{2007ARA&A..45..177C}
{Crowther}, P.~A. 2007, \araa, 45, 177

\bibitem[{{Cutri} {et~al.}(2012){Cutri}, {Wright}, {Conrow}, {Bauer},
  {Benford}, {Brandenburg}, {Dailey}, {Eisenhardt}, {Evans}, {Fajardo-Acosta},
  {Fowler}, {Gelino}, {Grillmair}, {Harbut}, {Hoffman}, {Jarrett},
  {Kirkpatrick}, {Leisawitz}, {Liu}, {Mainzer}, {Marsh}, {Masci}, {McCallon},
  {Padgett}, {Ressler}, {Royer}, {Skrutskie}, {Stanford}, {Wyatt}, {Tholen},
  {Tsai}, {Wachter}, {Wheelock}, {Yan}, {Alles}, {Beck}, {Grav}, {Masiero},
  {McCollum}, {McGehee}, {Papin}, \& {Wittman}}]{2012yCat.2311....0C}
{Cutri}, R.~M., {Wright}, E.~L., {Conrow}, T., {et~al.} 2012, VizieR Online
  Data Catalog, II/311

\bibitem[{{De Becker} {et~al.}(2017){De Becker}, {Benaglia}, {Romero}, \&
  {Peri}}]{2017A&A...600A..47D}
{De Becker}, M., {Benaglia}, P., {Romero}, G.~E., \& {Peri}, C.~S. 2017, \aap,
  600, A47

\bibitem[{{De Becker} \& {Raucq}(2013)}]{2013A&A...558A..28D}
{De Becker}, M. \& {Raucq}, F. 2013, \aap, 558, A28

\bibitem[{{Drew}(1990)}]{1990ASPC....7..230D}
{Drew}, J.~E. 1990, in Astronomical Society of the Pacific Conference Series,
  Vol.~7, Properties of Hot Luminous Stars, ed. C.~D. {Garmany}, 230--241

\bibitem[{{Eichler} \& {Usov}(1993)}]{1993ApJ...402..271E}
{Eichler}, D. \& {Usov}, V. 1993, \apj, 402, 271

\bibitem[{{Gooch}(1996)}]{1996ASPC..101...80G}
{Gooch}, R. 1996, in Astronomical Society of the Pacific Conference Series,
  Vol. 101, Astronomical Data Analysis Software and Systems V, ed. G.~H.
  {Jacoby} \& J.~{Barnes}, 80

\bibitem[{{Greisen}(2003)}]{2003ASSL..285..109G}
{Greisen}, E.~W. 2003, in Astrophysics and Space Science Library, Vol. 285,
  Information Handling in Astronomy - Historical Vistas, ed. A.~{Heck}, 109

\bibitem[{{Hamaguchi} {et~al.}(2018){Hamaguchi}, {Corcoran}, {Pittard},
  {Sharma}, {Takahashi}, {Russell}, {Grefenstette}, {Wik}, {Gull},
  {Richardson}, {Madura}, \& {Moffat}}]{2018NatAs...2..731H}
{Hamaguchi}, K., {Corcoran}, M.~F., {Pittard}, J.~M., {et~al.} 2018, Nature
  Astronomy, 2, 731

\bibitem[{{Intema}(2014)}]{2014ASInC..13..469I}
{Intema}, H.~T. 2014, in Astronomical Society of India Conference Series,
  Vol.~13, 469

\bibitem[{{Jones}(1985)}]{1985MNRAS.216..613J}
{Jones}, P.~A. 1985, \mnras, 216, 613

\bibitem[{{Lamberts} {et~al.}(2017){Lamberts}, {Millour}, {Liermann},
  {Dessart}, {Driebe}, {Duvert}, {Finsterle}, {Girault}, {Massi}, {Petrov},
  {Schmutz}, {Weigelt}, \& {Chesneau}}]{2017MNRAS.468.2655L}
{Lamberts}, A., {Millour}, F., {Liermann}, A., {et~al.} 2017, \mnras, 468, 2655

\bibitem[{{Leitherer} {et~al.}(1995){Leitherer}, {Chapman}, \&
  {Koribalski}}]{1995ApJ...450..289L}
{Leitherer}, C., {Chapman}, J.~M., \& {Koribalski}, B. 1995, \apj, 450, 289

\bibitem[{{Leitherer} {et~al.}(1997){Leitherer}, {Chapman}, \&
  {Koribalski}}]{1997ApJ...481..898L}
{Leitherer}, C., {Chapman}, J.~M., \& {Koribalski}, B. 1997, \apj, 481, 898

\bibitem[{{Leitherer} \& {Robert}(1991)}]{1991ApJ...377..629L}
{Leitherer}, C. \& {Robert}, C. 1991, \apj, 377, 629

\bibitem[{{Leser} {et~al.}(2017){Leser}, {Ohm}, {F{\"u}{\ss}ling}, {de
  Naurois}, {Egberts}, {Bordas}, {Klepser}, {Reimer}, {Reimer}, {Hinton}, \&
  {H.~E.~S.~S.~Collaboration}}]{Leser2017}
{Leser}, E., {Ohm}, S., {F{\"u}{\ss}ling}, M., {et~al.} 2017, International
  Cosmic Ray Conference, 35, 717

\bibitem[{{Martins} {et~al.}(2005){Martins}, {Schaerer}, \&
  {Hillier}}]{2005A&A...436.1049M}
{Martins}, F., {Schaerer}, D., \& {Hillier}, D.~J. 2005, \aap, 436, 1049

\bibitem[{{McMullin} {et~al.}(2007){McMullin}, {Waters}, {Schiebel}, {Young},
  \& {Golap}}]{2007ASPC..376..127M}
{McMullin}, J.~P., {Waters}, B., {Schiebel}, D., {Young}, W., \& {Golap}, K.
  2007, in Astronomical Society of the Pacific Conference Series, Vol. 376,
  Astronomical Data Analysis Software and Systems XVI, ed. R.~A. {Shaw},
  F.~{Hill}, \& D.~J. {Bell}, 127

\bibitem[{{Montes} {et~al.}(2011){Montes}, {Gonz{\'a}lez}, {Cant{\'o}}, {P{\'e
  }rez-Torres}, \& {Alberdi}}]{2011A&A...531A..52M}
{Montes}, G., {Gonz{\'a}lez}, R.~F., {Cant{\'o}}, J., {P{\'e }rez-Torres},
  M.~A., \& {Alberdi}, A. 2011, \aap, 531, A52

\bibitem[{{Montes} {et~al.}(2009){Montes}, {P{\'e}rez-Torres}, {Alberdi}, \&
  {Gonz{\'a}lez}}]{2009ApJ...705..899M}
{Montes}, G., {P{\'e}rez-Torres}, M.~A., {Alberdi}, A., \& {Gonz{\'a}lez},
  R.~F. 2009, \apj, 705, 899

\bibitem[{{Morton} \& {Wright}(1978)}]{1978MNRAS.182P..47M}
{Morton}, D.~C. \& {Wright}, A.~E. 1978, \mnras, 182, 47P

\bibitem[{{Muijres} {et~al.}(2012){Muijres}, {Vink}, {de Koter}, {M{\"u}ller},
  \& {Langer}}]{2012A&A...537A..37M}
{Muijres}, L.~E., {Vink}, J.~S., {de Koter}, A., {M{\"u}ller}, P.~E., \&
  {Langer}, N. 2012, \aap, 537, A37

\bibitem[{{Murphy} {et~al.}(2007){Murphy}, {Mauch}, {Green}, {Hunstead},
  {Piestrzynska}, {Kels}, \& {Sztajer}}]{2007yCat.8082....0M}
{Murphy}, T., {Mauch}, T., {Green}, A., {et~al.} 2007, VizieR Online Data
  Catalog, VIII/82

\bibitem[{{Murphy} {et~al.}(2009){Murphy}, {Sadler}, {Ekers}, {Massardi},
  {Hancock}, {Mahony}, {Ricci}, {Burke-Spolaor}, {Calabretta}, {Chhetri}, {de
  Zotti}, {Edwards}, {Ekers}, {Jackson}, {Kesteven}, {Lindley}, {Newton-McGee},
  {Phillips}, {Roberts}, {Sault}, {Staveley-Smith}, {Subrahmanyan}, {Walker},
  \& {Wilson}}]{2009yCat..74022403M}
{Murphy}, T., {Sadler}, E.~M., {Ekers}, R.~D., {et~al.} 2009, VizieR Online
  Data Catalog, J/MNRAS/402/2403

\bibitem[{{North} {et~al.}(2007){North}, {Davis}, {Tuthill}, {Tango}, \&
  {Robertson}}]{2007MNRAS.380.1276N}
{North}, J.~R., {Davis}, J., {Tuthill}, P.~G., {Tango}, W.~J., \& {Robertson},
  J.~G. 2007, \mnras, 380, 1276

\bibitem[{{Panagia} \& {Felli}(1975)}]{1975A&A....39....1P}
{Panagia}, N. \& {Felli}, M. 1975, \aap, 39, 1

\bibitem[{{Pittard}(2010)}]{2010MNRAS.403.1633P}
{Pittard}, J.~M. 2010, \mnras, 403, 1633

\bibitem[{{Pittard} \& {Dougherty}(2006)}]{2006MNRAS.372..801P}
{Pittard}, J.~M. \& {Dougherty}, S.~M. 2006, \mnras, 372, 801

\bibitem[{{Pshirkov}(2016)}]{2016MNRAS.457L..99P}
{Pshirkov}, M.~S. 2016, \mnras, 457, L99

\bibitem[{{Puls} {et~al.}(2008){Puls}, {Vink}, \&
  {Najarro}}]{2008A&ARv..16..209P}
{Puls}, J., {Vink}, J.~S., \& {Najarro}, F. 2008, \aapr, 16, 209

\bibitem[{{Reimer} {et~al.}(2006){Reimer}, {Pohl}, \&
  {Reimer}}]{2006ApJ...644.1118R}
{Reimer}, A., {Pohl}, M., \& {Reimer}, O. 2006, \apj, 644, 1118

\bibitem[{{Reitberger} {et~al.}(2017){Reitberger}, {Kissmann}, {Reimer}, \&
  {Reimer}}]{2017ApJ...847...40R}
{Reitberger}, K., {Kissmann}, R., {Reimer}, A., \& {Reimer}, O. 2017, \apj,
  847, 40

\bibitem[{{Reitberger} {et~al.}(2015){Reitberger}, {Reimer}, {Reimer}, \&
  {Takahashi}}]{2015A&A...577A.100R}
{Reitberger}, K., {Reimer}, A., {Reimer}, O., \& {Takahashi}, H. 2015, \aap,
  577, A100

\bibitem[{{Reynolds} {et~al.}(2012){Reynolds}, {Gaensler}, \&
  {Bocchino}}]{2012SSRv..166..231R}
{Reynolds}, S.~P., {Gaensler}, B.~M., \& {Bocchino}, F. 2012, \ssr, 166, 231

\bibitem[{{Rieger}(2017)}]{2017AIPC.1792b0008R}
{Rieger}, F.~M. 2017, in American Institute of Physics Conference Series, Vol.
  1792, 6th International Symposium on High Energy Gamma-Ray Astronomy, 020008

\bibitem[{{Rodr{\'\i}guez-Kamenetzky}
  {et~al.}(2016){Rodr{\'\i}guez-Kamenetzky}, {Carrasco-Gonz{\'a}lez}, {Araudo},
  {Torrelles}, {Anglada}, {Mart{\'\i}}, {Rodr{\'\i}guez}, \&
  {Valotto}}]{2016ApJ...818...27R}
{Rodr{\'\i}guez-Kamenetzky}, A., {Carrasco-Gonz{\'a}lez}, C., {Araudo}, A.,
  {et~al.} 2016, \apj, 818, 27

\bibitem[{{Romero} {et~al.}(1999){Romero}, {Benaglia}, \&
  {Torres}}]{1999A&A...348..868R}
{Romero}, G.~E., {Benaglia}, P., \& {Torres}, D.~F. 1999, \aap, 348, 868

\bibitem[{{Romero} {et~al.}(2003){Romero}, {Torres}, {Kaufman Bernad{\'o}}, \&
  {Mirabel}}]{2003A&A...410L...1R}
{Romero}, G.~E., {Torres}, D.~F., {Kaufman Bernad{\'o}}, M.~M., \& {Mirabel},
  I.~F. 2003, \aap, 410, L1

\bibitem[{{Runacres} \& {Owocki}(2002)}]{2002A&A...381.1015R}
{Runacres}, M.~C. \& {Owocki}, S.~P. 2002, \aap, 381, 1015

\bibitem[{{Rybicki} \& {Lightman}(1979)}]{1979rpa..book.....R}
{Rybicki}, G.~B. \& {Lightman}, A.~P. 1979, {Radiative processes in
  astrophysics}

\bibitem[{{Sault} {et~al.}(1995){Sault}, {Teuben}, \&
  {Wright}}]{1995ASPC...77..433S}
{Sault}, R.~J., {Teuben}, P.~J., \& {Wright}, M.~C.~H. 1995, in Astronomical
  Society of the Pacific Conference Series, Vol.~77, Astronomical Data Analysis
  Software and Systems IV, ed. R.~A. {Shaw}, H.~E. {Payne}, \& J.~J.~E.
  {Hayes}, 433

\bibitem[{{Tadhunter}(2016)}]{2016A&ARv..24...10T}
{Tadhunter}, C. 2016, Astronomy and Astrophysics Review, 24, 10

\bibitem[{{Tavani} {et~al.}(2009){Tavani}, {Sabatini}, {Pian}, {Bulgarelli},
  {Caraveo}, {Viotti}, {Corcoran}, {Giuliani}, {Pittori}, {Verrecchia},
  {Vercellone}, {Mereghetti}, {Argan}, {Barbiellini}, {Boffelli}, {Cattaneo},
  {Chen}, {Cocco}, {D'Ammando}, {Costa}, {DeParis}, {Del Monte}, {Di Cocco},
  {Donnarumma}, {Evangelista}, {Ferrari}, {Feroci}, {Fiorini}, {Froysland},
  {Fuschino}, {Galli}, {Gianotti}, {Labanti}, {Lapshov}, {Lazzarotto},
  {Lipari}, {Longo}, {Marisaldi}, {Mastropietro}, {Morelli}, {Moretti},
  {Morselli}, {Pacciani}, {Pellizzoni}, {Perotti}, {Piano}, {Picozza}, {Pilia},
  {Porrovecchio}, {Pucella}, {Prest}, {Rapisarda}, {Rappoldi}, {Rubini},
  {Soffitta}, {Trifoglio}, {Trois}, {Vallazza}, {Vittorini}, {Zambra},
  {Zanello}, {Santolamazza}, {Giommi}, {Colafrancesco}, {Antonelli}, \&
  {Salotti}}]{2009ApJ...698L.142T}
{Tavani}, M., {Sabatini}, S., {Pian}, E., {et~al.} 2009, \apj, 698, L142

\bibitem[{{van der Hucht} {et~al.}(2007){van der Hucht}, {Raassen}, {Mewe},
  {Antokhin}, {Rauw}, {Vreux}, {Schild}, \& {Schmutz}}]{2007ASPC..367..159V}
{van der Hucht}, K.~A., {Raassen}, A.~J.~J., {Mewe}, R., {et~al.} 2007, in
  Astronomical Society of the Pacific Conference Series, Vol. 367, Massive
  Stars in Interactive Binaries, ed. N.~{St. -Louis} \& A.~F.~J. {Moffat}, 159

\bibitem[{{White} \& {Chen}(1995)}]{1995IAUS..163..438W}
{White}, R.~L. \& {Chen}, W. 1995, in IAU Symposium, Vol. 163, Wolf-Rayet
  Stars: Binaries; Colliding Winds; Evolution, ed. K.~A. {van der Hucht} \&
  P.~M. {Williams}, 438

\bibitem[{{Wright} \& {Barlow}(1975)}]{1975MNRAS.170...41W}
{Wright}, A.~E. \& {Barlow}, M.~J. 1975, \mnras, 170, 41

\end{thebibliography}
%-------------------------------------------------------------------

\end{document}